\documentclass[journal,10pt]{IEEEtran}
\usepackage{graphicx,cite,amssymb,amsmath,psfrag,subfigure}
\usepackage{graphicx}
\usepackage{amssymb}
\usepackage{amsmath}
\usepackage{cite}
\usepackage{subfigure}
\usepackage{mathrsfs}
\usepackage[displaymath,mathlines]{lineno}
\usepackage{color}
\usepackage{tabulary}
\usepackage{multirow}
\usepackage[normalem]{ulem}
\usepackage{url}
\usepackage{hyperref}
\hypersetup{colorlinks,urlcolor=blue}

\hypersetup{colorlinks=false,pdfborder=000}
\usepackage{bm}

\newtheorem{theorem}{Theorem}

\newtheorem{corollary}{Corollary}
\newtheorem{proposition}{Proposition}

\newtheorem{remark}{Remark}

\newtheorem{proof}{Proof}

\newcommand{\mbs}[1]{\bm{#1}}
\newcommand{\mat}[1]{{\uppercase{\mbs{#1}}}}
\newcommand{\Id}{\mat{\mathrm{I}}}

\def\nn{\nonumber}
\renewcommand{\H}{{\scriptscriptstyle\mathsf{H}}}

\DeclareMathAlphabet{\mathpzc}{OT1}{pzc}{m}{it}

\DeclareMathOperator{\E}{\mathbb{E}} \DeclareMathOperator{\M}

\newcommand{\EX}[1]{\E\left\{{#1}\right\}}
\newcommand{\EXs}[2]{\E_{{#1}}\left\{{#2}\right\}}

\newcommand{\PDF}[2]{p_{{#1}}\left({#2}\right)}
\newcommand{\CG}[2]{\mathcal{CN}\left({#1},{#2}\right)}

\newcommand{\B}[1]{{\pmb{#1}}}

\newcommand{\Pu}{p_{r}}

\def\nn{\nonumber}
\renewcommand{\H}{{\scriptscriptstyle\mathsf{H}}}

\DeclareMathAlphabet{\mathpzc}{OT1}{pzc}{m}{it}

\makeatletter
\def\@setsize#1#2#3#4{
    \@nomath#1
    \let\@currsize#1
    \baselineskip #2
    \baselineskip \baselinestretch\baselineskip
    \parskip \baselinestretch\parskip
    \setbox\strutbox \hbox{
        \vrule height.7\baselineskip
            depth.3\baselineskip
            width\z@}
    \skip\footins \baselinestretch\skip\footins
    \normalbaselineskip\baselineskip#3#4}
\makeatother

\makeatletter
\newcommand{\setstretch}[1]{
    \def\baselinestretch{#1}%
    \@currsize
    }
\makeatother



\def\BibTeX{{\rm B\kern-.05em{\sc i\kern-.025em b}\kern-.08em
    T\kern-.1667em\lower.7ex\hbox{E}\kern-.125emX}}

\newcounter{eqncnt}

\newcounter{eqnback}

\begin{document}

\title{
    Performance of Massive MIMO Uplink with Zero-Forcing receivers under Delayed Channels}
\author{
       Anastasios K. Papazafeiropoulos, Hien Quoc  Ngo, and
        Tharm Ratnarajah
\thanks{Parts of this work were presented at the 2014 IEEE International Symposium on Personal, Indoor and Mobile Radio Communications (PIMRC)~\cite{Papazafeiropoulos6}.}

\thanks{
        A. K. Papazafeiropoulos was with Communications and Signal Processing Group, Imperial College London, London, U.K. Currently, he is  with the  Institute for Digital Communications (IDCOM), University of Edinburgh, Edinburgh, EH9 3JL, U.K., (email: a.papazafeiropoulos@ed.ac.uk). }
\thanks{
        H.~Q.\ Ngo is with Link\"{o}ping University, Link\"{o}ping, Sweden, and with Queen's University Belfast, Belfast, U.K.
        (email: hien.ngo@liu.se).
}
\thanks{
T. Ratnarajah is with Institute for Digital Communications
(IDCoM), University of Edinburgh, Edinburgh, U.K. (email:
t.ratnarajah@ed.ac.uk).}

 \thanks{This research was supported by a Marie Curie Intra-European Fellowship and ADEL project
 within the 7th European Community Framework Programme for Research of the European Commission under grant agreements no. [330806],
 IAWICOM and no. [619647], HARP. Also, this work was supported by the U.K. Engineering and Physical Sciences Research Council (EPSRC) under grant EP/L025299/1. The work of H.
Q. Ngo was supported by the Swedish Research Council (VR) and ELLIIT.}

}


\maketitle

\vspace{-0.5cm}
\begin{abstract}
In this paper, we analyze the  performance of the uplink communication of massive
multi-cell multiple-input multiple-output (MIMO) systems under the
effects of pilot contamination and delayed channels because of
terminal mobility. The base stations (BSs) estimate the channels through
the uplink training, and then use zero-forcing processing to
decode the transmit signals  from the users. The probability
density function (PDF) of the signal-to-interference-plus-noise
ratio is derived for any finite number of antennas. From this PDF,
we derive an achievable ergodic rate with a finite number of BS antennas in closed form. Insights of the impact of the
Doppler shift (due to terminal mobility) at the low
signal-to-noise ratio regimes are exposed. In addition, the effects on the outage probability are investigated. Furthermore, the power scaling law  and the
asymptotic performance result by infinitely increasing the 
numbers of antennas and terminals (while their ratio is fixed) are
provided. The numerical results demonstrate the performance loss
for various Doppler shifts. Among the interesting observations revealed is that
massive MIMO is favorable even in channel aging conditions.
\end{abstract}


\section{Introduction} \label{Sec:Introduction}

The rapidly increasing demand for wireless connectivity and
throughput is one of the motivations for the continuous evolution
of cellular networks \cite{Huawei,gesbert}. Massive multiple-input
multiple-output (MIMO) has been considered as a new promising
breakthrough technology due to its ability for achieving huge
spectral and energy efficiencies \cite{Marzetta,Rusek,Ngo_Energy,HLi}.
Its origin is found in \cite{Marzetta}, and it has been given many
alternative names such as very large multi-user MIMO, hyper-MIMO,
or full-dimension MIMO systems. In the typical envisioned
architecture, each base station (BS)  with an array of hundreds or even
thousands of antennas, exploiting the key idea of multi-user MIMO,
coherently serves tens or hundreds of single-antenna terminals
simultaneously in the same frequency band, respectively. This
difference in the number of BS antennas $N$ and the
number of terminals $K$ per cell provides unprecedented spatial
degrees of freedom that leads to a high throughput, allowing, in addition, low-complexity linear signal processing techniques
and avoiding inter-user interference because of the (near)
Unfortunately, a fundamental degradation, known as pilot contamination, degrades the performance of massive MIMO systems. It emerges from the re-use of the pilot sequences in other cells, and results in  inter-cell interference, even when the number of antennas becomes very large.

In massive MIMO, zero-forcing (ZF) processing is preferable since
it has low complexity and its performance is very close to that of
maximum-likelihood multi-user decoder and ``dirty paper coding''
\cite{Wang}. Plenty of research is  dedicated to  single-cell
networks with ZF receivers \cite{Caire}, and also to multi-cell
systems with perfect channel
state information (CSI) \cite{Mathaiou:ZF_receivers} and with the arising  pilot contamination~\cite{Jose}.

Despite that the theory of massive MIMO has been now well
established (see \cite{Marzetta} and references therein), the
impact of channel aging coming from  the relative movement of
terminals on  massive MIMO systems lacks
investigation in the literature. Channel aging problem occurs in
practical scenarios, e.g. in an urban area, where the mobility of
terminals is high. The fundamental challenge in these environments is how
to estimate the channel efficiently. To model the impact of
terminal mobility, a stationary ergodic Gauss-Markov block fading
\cite{Baddour,Truong,Papazafeiropoulos2,Papazafeiropoulos1,Papazafeiropoulos3,Papazafeiropoulos4}
is often used. With this channel, an autoregressive model is
associated with the Jakes' autocorrelation function which represents
the channel time variation.

It is known  that the  channel estimation overhead is  independent of the number of BS antennas, but it is  proportional to the number of terminals. The number of terminals that can be served  depends on the length of each frame which is designed based on the user mobility. Specifically, the higher the mobility of terminals is, the smaller the frame length is, which in turn degrades the system performance due to pilot contamination  (since fewer pilots can be sent). Furthermore, an increase of the user mobility  hampers the quality of channel estimation because the time variation of the channel increases, and thus, the constructed decoders and precoders lack precision.
Driven by these observations, this paper investigates the
robustness of massive MIMO against the practical setting of
terminal mobility that 
results in delayed and degraded CSI at the BS, and thus, imperfect
CSI. Such consideration is notably important because it can
provide the quantification of the performance loss in various
Doppler shifts. A limited effort for studying the time variation
of the channel because of the relative movement of terminals has taken place in~\cite{Truong}, where the authors provided
deterministic equivalents (DEs)\footnote{The deterministic
equivalents are deterministic tight approximations of functionals
of random matrices of finite size. Note that these approximations
are asymptotically accurate as the matrix dimensions grow to
infinity, but can be precise for small dimensions.} for the
maximal-ratio-combining (MRC) receivers in the uplink and the
maximal-ratio-transmission (MRT) precoders in the downlink. This
analysis was extended
in~\cite{Papazafeiropoulos1,Papazafeiropoulos2} by deriving DEs
for the minimum mean-square error (MMSE) receivers (for the
uplink) and regularized zero-forcing (for the downlink). In this
paper, extending~\cite{Papazafeiropoulos6}, we elaborate further on a generalized massive MIMO system uplink. Based
on the aforementioned literature, we propose a tractable model
that encompasses ZF receivers and describes the impact of terminal
mobility in a multicell system with an arbitrary number of BS antennas and terminals, which distinguishes it from previous works. The following are the main contributions of
this paper:
\begin{itemize}
\item 
In contrast to  \cite{Mathaiou:ZF_receivers}, which assumes perfect CSI, we consider more
practical settings where the channel is imperfectly estimated at
the BS. The effects of pilot contamination and channel
time variation are taken into account. The extension is not
straightforward because apart from the development of the model, the
mathematical manipulations are more difficult. Apart of this, the
results are contributory and novel.

\item We derive the probability density function (PDF) of the
signal-to-interference-plus-noise ratio (SINR), the corresponding
ergodic rate, and the outage probability for any finite number of
antennas in closed forms. For the sake of completeness, the link
of these results with previously known results is mentioned.
Furthermore, a simpler and more tractable lower bound for the
achievable uplink rate is derived.

\item We elaborate on the low signal-to-noise ratio (SNR) regime,
in order to get additional insights into the impact of Doppler
shift. In particular, we study the behaviors of the minimum
normalized energy per information bit to reliably convey any
positive rate and the wideband slope.


\item We evaluate the asymptotic performance for the case where
the number of BS antennas $N\to \infty$ and  for the case where both the number of BS antennas $N$ and
the number of the terminals $K$ go to infinity. This analysis aims
at providing accurate approximation results that replace the need
for lengthy Monte Carlo simulations.
\end{itemize}

Note that, although all the results incur significant mathematical
challenges, they can be easily evaluated. Moreover, the motivation behind the use of  DEs is to provide  deterministic tight
approximations, in order to avoid lengthy Monte-Carlo simulations.

The rest of this paper is structured as follows: Section~\ref{sec: system} presents the system model for the uplink of cellular systems with ZF receivers. In Section~\ref{Sec:Rate}, we provide the main results regarding the achievable uplink rate, its simple tight lower bound, and the outage probability. Furthermore, we shed light on the low-SNR regime of the system, and we investigate both the large number of antennas and large system (large number of antennas and users) limits.  The numerical results are discussed in Section~\ref{Numerical}, while Section~\ref{Conclusions} summarizes the paper.

\emph{Notation:} For matrices and vectors, we use boldface uppercase  and lowercase letters, respectively. The notations
$(\cdot)^\H$ and $(\cdot)^{\dagger}$ stand for the conjugate
transpose and the pseudo-inverse of a matrix as well as the Euclidean norm
of a vector is denoted by $\|\cdot\|$. The notation $x
\overset{\tt d}{\sim} y$ is used to denote that $x$ and $y$ have the same
distribution. Finally, we use $
\textcolor[rgb]{0.00,0.00,1.00}{\mathbf{z}} \sim
\CG{\mathbf{0}}{\B{\Sigma}}$ to denote a circularly symmetric
complex Gaussian vector 
\textcolor[rgb]{0.00,0.00,1.00}{$\mathbf{z}$} with zero mean and covariance
matrix $\B{\Sigma}$.

\setcounter{eqnback}{\value{equation}} \setcounter{equation}{12}
\begin{figure*}[!t]
\begin{align}\label{eq: SINR}
     \gamma_k
   \! =\!
        \frac{
            \alpha^2\Pu
            }{
            \alpha^2\Pu
            \sum_{i \neq l}^{L}\!
                \left\|\!
                    \left[\hat{\B{G}}_{ll}^{\dagger}[n\!-\!1]\right]_k
                   \! \hat{\B{G}}_{li}[n\!-\!1]
                \right\|^2
            \!+\!
        \Pu
            \sum_{i = l}^{L}\!
                \left\|\!
                    \left[\hat{\B{G}}_{ll}^{\dagger}[n\!-\!1]\right]_k
                   \! \tilde{\B{E}}_{li}[n]
                \!\right\|^2\!+\!
                 \left\|\!
                    \left[\hat{\B{G}}_{ll}^{\dagger}[n\!-\!1]\right]_k
                                   \! \right\|^2
            }.
\end{align}
\hrulefill
\end{figure*}
\setcounter{eqncnt}{\value{equation}}
\setcounter{equation}{\value{eqnback}}
\section{System Model} \label{sec: system}
We focus on a cellular network which has $L$ cells. Each cell
includes one $N$-antenna BS and $K$ single-antenna
terminals. We elaborate on the uplink transmission. The model is based
on the assumptions that: i) $N\geq K$, and ii) all terminals in
$L$ cells share the same time-frequency resource.
 Furthermore, we 
hypothesize  that the channels   change from symbol to symbol
under the channel aging impact~\cite{Truong} 
(we will discuss the channel  aging model later).

Denote by $\B g_{lik}[n]\in\mathbb{C}^{N\times 1}$ the channel
vector  between the $l$th BS and the $k$th terminal in
the $i$th cell at the $n$th symbol.  The channel  $\B
g_{lik}[n]\in\mathbb{C}^{N\times 1}$ is modeled by large-scale
fading (path loss and shadowing) and  small-scale fading as
follows:
\begin{align}
\B g_{lik}[n]= \sqrt{\beta_{lik}} \B h_{lik}[n],
\end{align}
where $\beta_{lik}$ represents  large-scale fading, and $\B
h_{lik}\in\mathbb{C}^{N\times 1}$ is the small-scale fading vector
between the $l$th BS and the $k$th terminal in the $i$th
cell with $\B h_{lik} \sim \CG{\B{0}}{\Id_N}$.

Let $\sqrt{p_{r}} \B{x}_i[n]\in \mathbb{C}^{K\times 1}$ be the
vector of transmit signals  from the $K$ terminals in the $i$th
cell at time instance $n$ ($p_{r}$ is the average transmit
power of each terminal. Elements of $\B{x}_i[n]$ are assumed to be
i.i.d. zero-mean and unit variance random variables (RVs). Then,
the $N\times1$ received signal vector at the $l$th BS is
\begin{align}\label{eq:rx1}
    \B{y}_l[n]
    =
        \sqrt{p_{r}}
        \sum_{i=1}^{L} \B{G}_{li}[n]  \B{x}_i[n]   +  \B{z}_l[n],~~~ l=1, 2, ...,
        L,
\end{align}
where $\B G_{li}[n]\triangleq \left[\B g_{li1}[n],\ldots, \B
g_{liK}[n]\right]\in \mathbb{C}^{N\times K}$ denotes the channel
matrix between the $l$th BS and the $K$ terminals in the
$i$th cell, and $\B{z}_l[n]\sim \CG{\B{0}}{\Id_N}$ is  the noise
vector at the $l$th BS.

\subsection{Uplink Training}
To coherently detect the transmit signals  from the $K$
terminals in the $l$th cell, the BS needs CSI knowledge.
Typically, the $l$th BS can estimate the channels from
the uplink pilots. During the training phase, we assume that the channel does not change, neglecting an error due to the channel aging effect \cite{Truong}. In general, this
assumption is not practical, but it yields a simple model which
enables us to analyze the system performance and to obtain initial
insights on the impact of channel aging. It is shown in \cite{Pohl} that, in most cases, the additional error term in the channel estimate which comes from the channel aging effect during the training phase can be neglected. In some cases, the investigation of the effect of this error is interesting. However, it yields a complicated model which is intractable for analysis. We leave this investigation for future work.

In the training phase, $K$ terminals in each cell are assigned
$K$ orthogonal pilot sequences, each has a length of $\tau$
symbols (it requires $\tau\geq K$). Owing to the limitation of the 
frame length\footnote{
The time-frequency resources are divided into frames of length $T$ symbols. In each frame, there are two basic phases: training phase and payload data transmission phase.}, the pilot sequences of all terminals in all
cells cannot be pairwisely orthogonal. We assume that the
orthogonal pilot sequences are reused from cell to cell (i.e., all
$L$ cells use the same set of $K$ orthogonal pilot sequences). 
As a result,  pilot contamination occurs \cite{Marzetta}. Let
$\B{\Psi}\in\mathbb{C}^{K\times \tau}$ be the pilot matrix
transmitted from the $K$ terminals in each cell, where the $k$th
row of $\B{\Psi}$ is the pilot sequence assigned for the $k$th
terminal. The matrix $\B{\Psi}$ satisfies $\B{\Psi}
\B{\Psi}^\H=\Id_K$. Then, the $N \times \tau$ received pilot
signal at the $l$th BS is given by
\begin{align}\label{eq MU-MIMO training}
\B{Y}^{\mathrm{tr}}_l[n]
    =        \sqrt{p_{\mathrm{tr}}}
        \sum_{i=1}^{L} \B{G}_{li}[n]\B {\Psi}   +  \B{Z}^{\mathrm{tr}}_l[n],~~~ l=1, 2, ..., L,
\end{align}
where  the superscript and subscript ``$\mathrm{tr}$'' imply the
uplink training, ${p_{\mathrm{tr}}}\triangleq \tau {p_{r}}$, and
$\B{Z}_l^{\mathrm{tr}}[n]\in \mathbb{C}^{N\times \tau}$ is the
additive noise. We consider that the elements of 
$\B{Z}^{\mathrm{tr}}_l[n]$ are   i.i.d.  $\CG{0}{1}$ RVs. With MMSE
channel estimation scheme, the estimate of ${\B g}_{lik}[n]$ is
\cite{Ngo_Energy}
\begin{align} \label{eq:MMSEchannelEstimate}
\hat{\B g}_{lik}[n] = & \beta_{lik} \B Q_{lk} \left( \sum_{j =1}^L
\B g_{ljk}[n] + \frac{1}{\sqrt{p_{\mathrm{tr}}}}\tilde{\B
z}^{\mathrm{tr}}_{lk}[n] \right)\!\!,
\end{align}
where $\B Q_{lk}\! \triangleq \!\left(\frac{1}{p_\mathrm{tr}}\!
+\! \sum_{i=1}^L \beta_{lik}\right)^{\!-1}\!\Id_N$, and $\tilde{\B
z}^{\mathrm{tr}}_{lk}[n] \!\sim\! \mathcal{CN}\!\left( \B
0,\Id_{N} \right)$ represents the noise which is independent of
$\B g_{ljk}[n]$.  Let $\hat{\B G}_{li}[n]\triangleq\left[\hat{\B
g}_{li1}[n],\ldots, \hat{\B g}_{liK}[n]\right]\in
\mathbb{C}^{N\times K}$. Then, $\hat{\B G}_{li}[n]$ can be
given by 
\begin{align}\label{interchannelEstimated}
 \hat{\B G}_{li}[n]= \hat{\B G}_{ll}[n] \B {{D}}_{i},
\end{align}
where $\B
{{D}}_{i}=\mathrm{diag}\big\{\frac{{\beta}_{li1}}{{\beta}_{ll1}},\frac{{\beta}_{li2}}{{\beta}_{ll2}},\ldots,\frac{{\beta}_{liK}}{{\beta}_{llK}}
\big\}$.

From the property of MMSE channel estimation, the  channel estimation error
and  the channel estimate  are independent. Thus, $\B
g_{lik}[n]$ can be rewritten as:
\begin{align}
\B g_{lik}[n] = \hat{\B g}_{lik}[n] + \tilde{\B
g}_{lik}[n],\label{eq:MMSEorthogonality}
\end{align}
where  $\tilde{\B g}_{lik}[n]$ and
 $\hat{\B g}_{lik}[n]$ are the independent  channel estimation error and channel estimate,
respectively. Furthermore, we  have  $\tilde{\B g}_{lik}[n] \sim
\CG{\B 0}{\left( \beta_{lik}-\hat{\beta}_{lik} \right)\B{\mathrm{
I}}_N}$ and
 $\hat{\B g}_{lik}[n] \sim \CG{\B 0}{\hat{\beta}_{lik}}$, where
$\hat{\beta}_{lik}\triangleq\frac{\beta^{2}_{lik}}{\sum_{j=1}^{L}\beta_{ljk}+1/p_{\mathrm{tr}}}$.
Here we assume that   $\beta_{lik}$, $\hat{\beta}_{lik}$, and $\B
Q_{lk}$ are independent of $n$ $\forall l$, $i$,  and $k$. This
assumption is reasonable since these values depend on large-scale
fading which changes very slowly with time.

\subsection{Delayed Channel Model}
Besides pilot contamination, in any common propagation scenario, a
relative movement takes place between the antennas and the
scatterers that degrades more channel's  performance. Under these
circumstances, the channel is time-varying and needs to be modeled
by the famous Gauss-Markov block fading model, which is basically
an autoregressive model of certain order that incorporate
two-dimensional isotropic scattering (Jakes model). More
specifically, our analysis achieves to express the current channel
state in terms of its past samples. For the sake of  tractable analytical and computational
simplicity, we focus on the following simplified autoregressive
model of order $1$~\cite{Baddour,Truong,Papazafeiropoulos2,Papazafeiropoulos1,Papazafeiropoulos3,Papazafeiropoulos4}
\begin{align}\label{eq:aut}
 \B g_{lik}[n]=\alpha {\B g}_{lik}[n-1]+\B e_{lik}[n],
\end{align}
where  $\B e_{lik}[n]\sim \CG{\B{0}}{\left( 1-\alpha^2
\right)\beta_{lik}\B{\mathrm{ I}}_N}$ is the stationary Gaussian
channel error vector because of the time variation of the channel,
independent of ${\B g}_{lik}[n-1]$. In \eqref{eq:aut}, $\alpha$ is
the temporal correlation parameter, given by
\begin{align}\label{eq:autalpha}
\alpha\!=\!\mathrm{J}_{0}\left( 2 \pi f_{D} T_{s} \right),
\end{align}
where $\mathrm{J}_{0}(\cdot)$ is the zeroth-order Bessel function
of the first kind, $f_{D}$ is the maximum Doppler shift, and
$T_{s}$ is the channel sampling period.  The maximum Doppler shift
$f_{D}$ is equal to $\frac{v f_{c}}{c}$, where $v$ is the relative
velocity of the terminal, $c$ is the speed of light, and $f_{c}$
is the carrier frequency.\footnote{The following analysis holds for any maximum Doppler shift obeying to Eq.~\eqref{eq:autalpha}. For a given temporal
correlation parameter $\alpha$, $f_{D}$ can be evaluated numerically. } It is assumed that  $\alpha$ is accurately obtained at the
BS via a rate-limited backhaul link.

Plugging \eqref{eq:MMSEorthogonality} into \eqref{eq:aut}, we
obtain a model which represents both effects of channel estimation
error due to pilot contamination and channel aging:
\begin{align}
 \B g_{lik}[n]&=\alpha {\B g}_{lik}[n-1]+\B e_{lik}[n]\nonumber\\
&=\alpha \hat{\B g}_{lik}[n-1]+ \tilde{\B
e}_{lik}[n],\label{eq:MMSEchannelEstimate}
\end{align}
where $\tilde{\B {e}}_{lik}[n]\triangleq \alpha \tilde{\B
g}_{lik}[n-1]+\B e_{lik}[n]\sim \CG{\B{0}}{\left(
\beta_{lik}-\alpha^{2}\hat{\beta}_{lik} \right)\B{\mathrm{ I}}_N}$
is independent of $\hat{\B g}_{lik}[n-1]$.

\subsection{Zero-Forcing Receiver}
Substituting \eqref{eq:MMSEchannelEstimate} into \eqref{eq:rx1},
the received signal at the $l$th BS can be rewritten as
\begin{align}\label{eq MU-MIMO 1}
    \B{y}_l[n]
    \!=\!
        \alpha \sqrt{p_{r}}
        \sum_{i=1}^{L}\!\B{\hat{G}}_{li}[n\!-\!1]  \B{x}_i[n]   +  \sqrt{p_{r}} \sum_{i=1}^{L}\! \B{\tilde{E}}_{li}[n]  \B{x}_i[n]  \!+\!
        \B{z}_l[n],
\end{align}
where $\tilde{\B E}_{li}\triangleq\left[\tilde{\B
e}_{li1}[n],\ldots, \tilde{\B e}_{liK}[n]\right]\in
\mathbb{C}^{N\times K}$.  With ZF processing, the received signal
$\B{y}_l[n]$ is first multiplied with
$\alpha^{-1}\hat{\B{G}}_{ll}^\dagger[n-1]$ as follows:
\begin{align}
 \label{eq MU-MIMO 2}
\! &\!\B{r}_l[n]
    =
       \sqrt{p_{r}}\B{x}_l[n]+
         \sqrt{p_{r}} \sum_{i\ne l}^{L} \hat{\B{G}}_{ll}^\dagger[n-1] \B{\hat{G}}_{li}[n-1]  \B{x}_i[n]            \nn\\
  \!  \!\!\! \! \!  \!\! &\!+\!  \alpha^{\!-\!1}\!\sqrt{p_{r}} \sum_{i=1}^{L}\! \hat{\B{G}}_{ll}^\dagger [n\!-\!1] \B{\tilde{E}}_{li}[n]  \B{x}_i[n] \!+\!\alpha^{\!-\!1}\hat{\B{G}}_{ll}^\dagger[n\!-\!1] \B{z}_l[n].
\end{align}

Then, the $k$th element of $\B{r}_l[n]$ is used to decode the transmit
signal  from the $k$th terminal, ${x}_{lk}[n]$. The
$k$th element of $\B{r}_l[n]$ is
\begin{flalign}
 \label{eq MU-MIMO 2}
  \!\!  &\!\!\!{r}_{lk}[n]
    =
         \sqrt{p_{r}}{x}_{lk}[n]\!+\!
         \sqrt{p_{r}}\! \sum_{i\ne l}^{L}\! \left[\hat{\B{G}}_{ll}^\dagger[n-1]\right]_{k}\! \B{\hat{G}}_{li}[n\!-\!1]  \B{x}_i[n]  \nn\\
          \!\!\!\! &\!\!\!+ \! \frac{1}{\alpha}\!\sqrt{p_{r}} \!\!\sum_{i=1}^{L} \!\left[\hat{\B{G}}_{ll}^\dagger[n\!-\!1]\right]_{k}\!\!\B{\tilde{E}}_{li}[n] \B{x}_i[n]  \!+\!\frac{1}{\alpha}\!\!\left[\hat{\B{G}}_{ll}^\dagger[n\!-\!1]\right]_{k}\! \B{z}_l[n],   \end{flalign}
where  $\left[\B A\right]_{k}$ denotes the $k$th row of matrix $\B
A$, and ${x}_{lk}[n]$ is the $k$th element of $\B{x}_{l}[n]$. 
By
treating \eqref{eq
MU-MIMO 2} as a single-input single-output (SISO) system, we
obtain the SINR of the transmission from the $k$th user in the
$l$th cell to its BS given by \eqref{eq: SINR} shown at the top of the previous
page.
Henceforth, we assume that this SINR is obtained under the assumption that the
$l$th BS does not need the instantaneous knowledge of the terms in the denominator of \eqref{eq: SINR}, but only of their statistics, which can be easily acquired, especially, if  they change over a long-time scale. More specifically, the BS knows the probability distribution of the actual channel given the available estimate, i.e., if we denote the probability $p$, we have $P_{\B{G}|\hat{\B{G}}}=p_{\tilde{\B{E}}}=p_{(\B{G}-\hat{\B{G}})}$. 

\setcounter{eqnback}{\value{equation}} \setcounter{equation}{17}
\begin{figure*}[!t]
\begin{align}
    \mathcal{J}_{m,n}\left( a,b,\alpha \right)
    &\triangleq
        \sum_{r=0}^{m}
            \binom{m}{r}
            \left(-b\right)^{m-r}
        \left[
            \sum_{s=0}^{n+r}
            \frac{
                \left(n+r \right)^s
                b^{n+r-s}
                }{
                \alpha^{s+1} a^{m-s}
                }
            \mathrm{Ei} \left(\!-b\right)
            -
            \frac{
                \left(n+r \right)^{n+r}
                e^{\alpha b/a}
                }{
                \alpha^{n+r+1}
                a^{m-n-r}
                }
            \mathrm{Ei} \left(\!-\frac{\alpha b}{a}-b \!\right)
        \right.
    \nonumber
    \\
    & \hspace{4.5 cm}
        \left.
        +
        \frac{e^{-b}}{\alpha }
        \sum_{s=0}^{n+r-1}
        \sum_{u=0}^{n+r-s-1}
        \frac{
            u!
            \left(n+r\right)^s \binom{n+r-s-1}{u}
            b^{n+r-s-u-1}
            }{
            \alpha^s
            a^{m-s}
            \left(\alpha/a + 1 \right)^{s+1}
            }
        \right].\label{I Func}
\end{align}
\hrulefill
\end{figure*}
\setcounter{eqncnt}{\value{equation}}
\setcounter{equation}{\value{eqnback}}
\section{Achievable Uplink Rate} \label{Sec:Rate}
This section provides the achievable rate analysis for finite and
infinite number of BS antennas by accomodating the
 effects of pilot contamination and channel aging.

\subsection{Finite-$N$ Analysis}\label{Sec:Rate_CF}

Denote by $\mathbf{\mathcal{A}}_k\triangleq \mathrm{diag} \left(
\tilde{\B{D}}_{l1},\ldots,\tilde{\B{D}}_{lL} \right)$, where
$\tilde{\B D}_{li}$ a $K \times K$ diagonal matrix whose $k$th
diagonal element is $\left[\tilde{\B D}_{li}\right]_{kk}=\left(
\beta_{lik}-\alpha^{2}\hat{\beta}_{lik} \right)$. Then the
distribution of the SINR for the uplink transmission from the
$k$th terminal is given in the following proposition.
 \addtocounter{equation}{1}

\begin{proposition}\label{Prop 1}
The SINR of transmission from the $k$th terminal in the $l$th cell
to its BS, under the delayed channels, is distributed as
\begin{align} \label{eq Prop1 1}
    \gamma_k
    \mathop \sim \limits^{\tt d}
                \frac{
                    \alpha^2 p_{r} X_k[n-1]
                    }{
                    \alpha^2p_{r} C X_k[n-1] +p_{r} Y_k[n]+ 1
                    },
\end{align}
where $C \triangleq \sum_{i \ne
l}^{L}\left(\frac{\beta_{lik}}{\beta_{llk}} \right)^2$  is a
deterministic constant, $X_k$ and $Y_k$ are independent RVs whose
PDFs are, respectively, given by
\begin{align} \label{eq PDF1 1}
   \!\!\PDF{X_k}{x}
   &\!=\!
        \frac{
            e^{-x/\hat{ \beta}_{llk}}
            }{
            \left(N-K\right)!
           \hat{ \beta}_{llk}
            }
        \left(
            \frac{
                x
                }{
               \hat{ \beta}_{llk}
                }
        \right)^{N-K}, ~ x \geq 0,
    \\
    \!\!\PDF{Y_k}{y}
    &\!=\!\!
        \sum_{p=1}^{\varrho\left(\! \mathbf{\mathcal{A}}_k\!\right)}\!
        \sum_{q=1}^{\tau_p \left(\! \mathbf{\mathcal{A}}_k\!\right)} \!\!
            \!\!\mathcal{X}_{p,q} \left(\! \mathbf{\mathcal{A}}_k\!\right)
           \! \frac{
                \mu_{k,p}^{-q}
            }{
                \left(
                    q\!-\!1
                \right)
                !
            }
            y^{q-1}
            e^{\frac{-y}{\mu_{k,p}}}
      ,
         ~
        y \geq 0.\label{eq PDF1 3}
\end{align}
In \eqref{eq PDF1 3}, $\mathcal{X}_{p,q} \left(
\mathbf{\mathcal{A}}_k\right)$ is the $\left(p,q\right)$th
characteristic coefficients of $\mathbf{\mathcal{A}}_k$, defined
in \cite[Definition~4]{SW:08:IT};  $\varrho\left(
\mathbf{\mathcal{A}}_k\right)$ is the numbers of distinct diagonal
elements of $\mathbf{\mathcal{A}}_k$; $\mu_{k,1}, ...,
\mu_{k,\varrho\left( \mathbf{\mathcal{A}}_k\right)}$ are the
distinct diagonal elements of $\mathbf{\mathcal{A}}_k$ in
decreasing order; and $\tau_p \left(
\mathbf{\mathcal{A}}_k\right)$ are the multiplicities of
$\mu_{k,p}$.
\begin{proof}
See Appendix~\ref{appproofPro1}.
\end{proof}
\end{proposition}


 \begin{remark}
Following the behavior of the Bessel function
$\mathrm{J}_{0}(\cdot)$, the SINR presents ripples with zero and
peak points with respect to the 
relative velocity of the user with respect to the BS. In the extreme
case of $\alpha=1$  (corresponding to the case where there is no
relative movement of the terminal), \eqref{eq Prop1 1} represents
the result for the case of without channel aging impact. In
another extreme case where $\alpha \to 0$ (i.e. velocity is very
high), SINR becomes zero. Furthermore, if we assume no time
variation and the training intervals can be long enough so that
all pilot sequences are orthogonal, our result coincides with
\cite[Eq.~(6)]{Mathaiou:ZF_receivers}.
 \end{remark}

 \begin{corollary}\label{coll1}
When the uplink power grows large, the SINR $\gamma_k$ is bounded:
\begin{align} \label{pu limit}
 \gamma_k\big|_{p_{r} \rightarrow \infty}
    \mathop \sim \limits^{\tt d}
            \frac{
            \alpha^2  X_k[n-1]
            }{
            \alpha^2 C X_k[n-1]
                 +Y_k[n]}.
\end{align}
Corollary~\ref{coll1} brings  an important insight on the system
performance, when $p_{r}$ is large. As seen in~\eqref{pu limit},
there is a finite SINR ceiling when $p_{r}\to\infty$, which emerges because of the simultaneous increases of the desired signal power and the
interference powers when $p_{r}$ increases.
\end{corollary}

Having obtained the PDF of the SINR, and by defining the function
$\mathcal{J}_{m,n}\left( a,b,\alpha \right)$ as in \eqref{I Func}
shown at the top of the next page, where
$\mathrm{Ei}\left(\cdot\right)$ denotes the exponential integral
function \cite[Eq.~(8.211.1)]{GR:07:Book}, we first obtain the
exact $R_{lk}\left( p_{r},\alpha \right)$ and a simpler lower
bound $R_{L}\left( p_{r},\alpha \right)$ as follows:
 \addtocounter{equation}{1}

\begin{theorem} \label{Theo 1}
The uplink ergodic achievable rate  of transmission from the $k$th
terminal in the $l$th cell to its BS for any finite
number of antennas, under delayed channels, is
\begin{align}
   \label{Rate prop}
\!\!R_{lk}\!\left( p_{r},\alpha \right)
    \!=\!\!\!
\sum_{p=1}^{\varrho\left(\! \mathbf{\mathcal{A}}_k\!\right)}\!
        \sum_{q=1}^{\tau_p \left(\! \mathbf{\mathcal{A}}_k\!\right)}\!
           \!\!\! \frac{
                 \mathcal{X}_{p,q} \left(\! \mathbf{\mathcal{A}}_k\!\right)
               \mu_{k,p}^{-q} \log_2 e
            }{
                \left(
                    q\!-\!1
                \right)
        !\!
                \left(N\!-\!K\right)!
                \hat{\beta}_{llk}^{N\!-\!K\!+\!1}\!
            }    \left(  \mathcal{I}_{1}\!-\!\mathcal{I}_{2} \right)\!,
\end{align}
where $\mathcal{I}_{1}$ and $\mathcal{I}_{2}$ are given by
\eqref{eq Prop2 1a} and \eqref{eq Prop2 1b} shown at the top of
the next page, and where $U\left(\cdot,\cdot,\cdot \right)$ is the
confluent hypergeometric function of the second kind
\cite[Eq.~(9.210.2)]{GR:07:Book}.
\begin{proof}
See Appendix~\ref{sec:proof:prop rate1}.
\end{proof}
\end{theorem}

\setcounter{eqnback}{\value{equation}} \setcounter{equation}{19}
\begin{figure*}[!t]
\begin{align} \label{eq Prop2 1a}
      \mathcal{I}_{1}
    &\!\triangleq\! \sum_{t=0}^{N-K}\left[
            - e^{\frac{1}{\hat{\beta}_{llk} \alpha^2 p_{r} \left( C+1 \right)}}
            \mathcal{J}_{q\!-\!1, N\!-\!K\!-\!t}\left(\!\frac{1}{\hat{\beta}_{llk}\alpha^{2}\left( C+1 \right)},\frac{1}{\hat{\beta}_{llk} \alpha^2 p_{r} \left( C+1 \right)},\frac{1}{\mu_{k,p}}\!-\!\frac{1}{\hat{\beta}_{llk}\alpha^{2}\left( C+1 \right)}\! \right)
        \right.\nn \\
                &+
                \sum_{u=1}^{N-K-t}
                \frac{\left(u-1 \right)! \left(-1\right)^{u} p_{r}^{-q}}{\left( {\hat{\beta}_{llk}\alpha^2 p_{r} \left( C+1 \right)}\right)^{N-K-t-u} }
                  \left.
        \Gamma\left(q\right)
                U\left(q, q+1+N-K-t-u, \frac{1}{\mu_{k,p}p_{r} }\right)
            \right],\\
                   \mathcal{I}_{2}
    &\!\triangleq\! \! \sum_{t=0}^{N-K}\left[
            - e^{\frac{1}{\hat{\beta}_{llk} \alpha^2 p_{r} C}}
            \mathcal{J}_{q\!-\!1, N\!-\!K\!-\!t}\left(\!\frac{1}{\hat{\beta}_{llk}\alpha^{2}C},\frac{1}{\hat{\beta}_{llk} \alpha^2 p_{r} C},\frac{1}{\mu_{k,p}}\!-\!\frac{1}{\hat{\beta}_{llk}\alpha^{2}C}\! \right)
        \right.\nn\\
        &        +
                \sum_{u=1}^{N-K-t}
                \frac{\left(u-1 \right)! \left(-1\right)^{u} p_{r}^{-q}}{\left( {\hat{\beta}_{llk}\alpha^2 p_{r} C}\right)^{N-K-t-u} }
         \left.
                 \   \Gamma\left(q\right)
                U\left(q, q+1+N-K-t-u, \frac{1}{\mu_{k,p}p_{r} }\right)
            \right]\label{eq Prop2 1b},
\end{align}
\hrulefill
\end{figure*}
\setcounter{eqncnt}{\value{equation}}
\setcounter{equation}{\value{eqnback}}

In the case that all diagonal elements of $\mathbf{\mathcal{A}}_k$
are distinct, we have $\varrho\left( \mathbf{\mathcal{A}}_k\right)
= KL$, $\tau_p \left( \mathbf{\mathcal{A}}_k\right)=1$, and
$\mathcal{X}_{p,1} \left( \mathbf{\mathcal{A}}_k\right)
    =    \prod_{q=1, q \neq p}^{KL}
    \left(
        1
        -
        \frac{\mu_{k,q}}{\mu_{k,p}}
    \right)^{-1}$. The uplink rate becomes
 \addtocounter{equation}{2}
\begin{align} \label{eq Prop2 1}
  R_{lk}\!\left(\! p_{r},\alpha \!\right)    &\!=\!
         \sum_{p=1}^{KL}
    \sum_{t=0}^{N-K}\!
             \frac{
                \prod_{q=1, q \neq p}^{KL}
    \left(\!
        1
        \!-\!
        \frac{\mu_{k,q}}{\mu_{k,p}}
    \!\right)^{-1}\!\!\!\log_2 e
                            }{
                \left(N\!-\!K\!-\!t\right)! (-1)^{N-K-t}\mu_{k,p}
                            }\left( \bar{\mathcal{I}}_1\!-\!\bar{\mathcal{I}}_2 \right),
               \end{align}
where $\bar{\mathcal{I}}_{1}$ and $\bar{\mathcal{I}}_{2}$ are
given by \eqref{eq Prop2 1aa} and \eqref{eq Prop2 1bb} shown at
the top of the next page. Note that, we have used the identity
$U\left(1, b, c \right)=e^{x}x^{1-b}\Gamma\left( b-1,x \right)$
\cite[Eq. (07.33.03.0014.01)]{Wolfram} to obtain \eqref{eq Prop2
1}.

\setcounter{eqnback}{\value{equation}} \setcounter{equation}{22}
\begin{figure*}[!t]
\begin{align} \label{eq Prop2 1aa}
      \bar{\mathcal{I}}_{1}
    &\!=\!\sum_{t=0}^{N-K}\left[
            - e^{\frac{1}{\hat{\beta}_{llk} \alpha^2 p_{r} \left( C+1 \right)}}
            \mathcal{J}_{0, N\!-\!K\!-\!t}\left(\!\frac{1}{\hat{\beta}_{llk}\alpha^{2}\left( C+1 \right)},\frac{1}{\hat{\beta}_{llk} \alpha^2 p_{r} \left( C+1 \right)},\frac{1}{\mu_{k,p}}\!-\!\frac{1}{\hat{\beta}_{llk}\alpha^{2}\left( C+1 \right)}\! \right)
        \right.\nn\\
                &+
                \sum_{u=1}^{N-K-t}
                \frac{\left(u-1 \right)! \left(-1\right)^{u} }{\left( {\hat{\beta}_{llk}\alpha^2  \left( C+1 \right)}\right)^{N-K-t-u} }e^{\frac{1}{\mu_{k,p}p_{r}}}\mu_{k,p}^{N+1-K-t-u}\Gamma\left(N+1-K-t-u, \frac{1}{\mu_{k,p}p_{r} }\right)\Biggl. \Biggr],\\
      \bar{\mathcal{I}}_{2}
    &\!=\! \! \sum_{t=0}^{N-K}\left[
            - e^{\frac{1}{\hat{\beta}_{llk} \alpha^2 p_{r} C}}
            \mathcal{J}_{1, N\!-\!K\!-\!t}\left(\!\frac{1}{\hat{\beta}_{llk}\alpha^{2}C},\frac{1}{\hat{\beta}_{llk} \alpha^2 p_{r} C},\frac{1}{\mu_{k,p}}\!-\!\frac{1}{\hat{\beta}_{llk}\alpha^{2}C}\! \right)
        \right.\nn\\
        &        +
                \sum_{u=1}^{N-K-t}
                \frac{\left(u-1 \right)! \left(-1\right)^{u} }{\left( {\hat{\beta}_{llk}\alpha^2  C}\right)^{N-K-t-u} }
         \left.
                    e^{\frac{1}{\mu_{k,p}p_{r}}}\mu_{k,p}^{N+1-K-t-u}\Gamma\left(N+1-K-t-u, \frac{1}{\mu_{k,p}p_{r} }\right)
            \right].\label{eq Prop2 1bb}
\end{align}
\hrulefill
\end{figure*}
\setcounter{eqncnt}{\value{equation}}
\setcounter{equation}{\value{eqnback}}

\setcounter{eqnback}{\value{equation}} \setcounter{equation}{26}
\begin{figure*}[!t]
\begin{align}
 \displaystyle P_{{\tt out}}\!\left( \gamma_{{\tt th}}\right)
 \!=\!
 \left\{\!\!\!
   \begin{array}{l}
     1, \quad \text{if $\gamma_{{\tt th}} \geq 1/C$} \\
      1\!-\!e^{-\frac{\gamma_{{\tt th}}}{\hat{\beta}_{llk}\left( \alpha^2 p_{r}-\alpha^{2} p_{r} C \gamma_{{\tt th}} \right)}}\!
\sum\limits_{p=1}^{\varrho\left( \mathbf{\mathcal{A}}_k\right)}\!
\sum\limits_{q=1}^{\tau_p \left( \mathbf{\mathcal{A}}_k\right)}\!
 \sum\limits_{t=0}^{N-K} \!\!
  \sum\limits_{s=0}^{t}\!\! \binom{t}{s}
 \mathcal{X}_{p,q} \!\left(
 \mathbf{\mathcal{A}}_k\!\right) \frac{\mu_{k,p}^{-q}}{\left(q\!-\!1\right)} \Gamma\left( s\!+\!q \right)\!\!\left( \hat{\beta}_{llk}\left( \alpha^{2}\!- \!\alpha^{2} C \gamma_{{\tt th}}\right) \!\!\right)^{s+q},  \quad \text{if $\gamma_{{\tt th}} < 1/C$}.\label{eq:out}\\
   \end{array}
 \right.
\end{align}
\hrulefill
\end{figure*}
\setcounter{eqncnt}{\value{equation}}
\setcounter{equation}{\value{eqnback}}
The achievable ergodic rate of the $k$th terminal in the $l$th
cell, given by~\eqref{eq Prop2 1}, is rather complicated. We next
proceed with the derivation of a lower bound. Indeed, the
following proposition provides a relatively simple analytical
expression for a lower of $R_{lk}$ which is very tight (see the
numerical result section).
\begin{proposition}\label{Prop BoundRate 1}
The uplink ergodic rate  from the $k$th terminal in the $l$th cell
to its BS, considering delayed channels, is lower
bounded by  $ R_{L}\left( p_{r},\alpha \right)$: \addtocounter{equation}{2}
\begin{align}\label{eq: LB}
 \!\! &R_{lk}\left( p_{r},\alpha \right)  \!\geq\! R_{L}\left( p_{r},\alpha
 \right)\nonumber\\
&\triangleq \log_2\!\!\left(\!\!1\!+ \!\frac{1}{C+\frac{1}{\left(
N-K \right)\alpha^{2} \hat{
\beta}_{llk}}\!\left(\sum\limits_{i=1}^{L}\sum\limits_{k=1}^{K}\!\!
\left(\! \beta_{lik}\!-\!\alpha^{2}\hat{\beta}_{lik}
\!\right)\!+\!\frac{1}{p_{r}}\! \right)}\!\!\right)\!.
\end{align}
\begin{proof}
See Appendix~\ref{propBound Rate}.
\end{proof}
\end{proposition}

According to~\eqref{eq: LB}, it can be easily seen that the slower
the channel varies (higher $\alpha$), the higher the lower bound
of $R_{lk}\left( p_{r},\alpha \right)$ is.

\subsubsection{Outage Probability}
In the case of block fading, the study of the outage probability
is of particular interest. Basically, it defines the probability
that the instantaneous SINR $\gamma_{k}$ falls below a given
threshold value $\gamma_{{\tt th}}$:
\begin{align}
 P_{{\tt out}}\left( \gamma_{{\tt th}}\right)
 &= {\tt Pr}\left( \gamma_k\leq\gamma_{{\tt th}}\right).
\end{align}
\begin{theorem}\label{theo 2}
The outage probability of transmission from the $k$th terminal in
the $l$th cell to its BS is  given by \eqref{eq:out}.
\begin{proof}
See Appendix~\ref{proof:Prop out}.
\end{proof}
\end{theorem}

The outage probability increases as the terminal mobility
increases, i.e., as $\alpha$ decreases.
Furthermore, when $p_r\to\infty$, the outage probability, given by \eqref{eq:out}, becomes as in \eqref{pout2}, shown at the top of the next page. Since $P_{{\tt out}}\left( \gamma_{{\tt th}}\right)$ is independent of $p_r$,  the diversity order is equal to zero.
\setcounter{eqnback}{\value{equation}} \setcounter{equation}{27}

\begin{figure*}
\begin{align}
 \displaystyle P_{{\tt out}}\!\left( \gamma_{{\tt th}}\right)
 \!\to\!
 \left\{\!\!\!
   \begin{array}{l}
     1, \quad \text{if $\gamma_{{\tt th}} \geq 1/C$} \\
      1\!-\!
\sum\limits_{p=1}^{\varrho\left( \mathbf{\mathcal{A}}_k\right)}\!
\sum\limits_{q=1}^{\tau_p \left( \mathbf{\mathcal{A}}_k\right)}\!
 \sum\limits_{t=0}^{N-K} \!\!
  \sum\limits_{s=0}^{t}\!\! \binom{t}{s}
 \mathcal{X}_{p,q} \!\left(
 \mathbf{\mathcal{A}}_k\!\right) \frac{\mu_{k,p}^{-q}}{\left(q\!-\!1\right)} \Gamma\left( s\!+\!q \right)\!\!\left( \hat{\beta}_{llk}\left( \alpha^{2}\!- \!\alpha^{2} C \gamma_{{\tt th}}\right) \!\!\right)^{s+q},  \quad \text{if $\gamma_{{\tt th}} < 1/C$}.\label{pout2}\\
    \end{array}
 \right.
\end{align} 
\hrulefill
\end{figure*}
\setcounter{eqnback}{\value{equation}} \setcounter{equation}{28}

\subsection{Characterization in the Low-SNR Regime}
Even though  Theorem~\ref{Theo 1} renders possible the exact
derivation of the achievable uplink rate, it appears deficient to
provide an insightful dependence on the various parameters such as
the number of BS antennas and the transmit power. On
that account, the study of the low power cornerstone, i.e., the
low-SNR regime, is of great significance. There is no reason to
consider the high-SNR regime, because in this regime an important
metric such as the high-SNR slope $\mathcal{S}_{\infty}=
\lim_{p_{r} \rightarrow 0} \frac{R_{lk}\left( p_{r}, \alpha
\right)}{\mathrm{log}_2 p_{r}}$ \cite{Tulino} is zero due to the
finite rate, as shown in \eqref{pu limit}.

\subsubsection{Low-SNR Regime}
In case of low-SNR, it is possible to represent the rate by means
of second-order Taylor approximation as\addtocounter{equation}{0}
\begin{align} R_{lk}\left( p_{r}, \alpha
\right)=\dot{R}_{lk}\left( 0, \alpha
\right)p_{r}+\ddot{R}_{lk}\left( 0, \alpha
\right)\frac{p_{r}^{2}}{2}+o\left( p_{r}^{2} \right),
\end{align}
where $\dot{R}_{lk}\left( p_{r}, \alpha \right)$ and
$\ddot{R}_{lk}\left( p_{r}, \alpha \right)$ denote the first and
second derivatives of $R_{lk}\left( p_{r}, \alpha \right)$ with
respect to SNR $p_{r}$. In fact, these parameters enable us to
examine the energy efficiency in the regime of low-SNR by means of
two key element parameters, namely the minimum transmit energy per
information bit, $\frac{E_{b}}{N_{0}{_{\mathrm{min}}}}$, and the
wideband slope $S_{0}$~\cite{Verdu}. Especially, we have
\begin{align}\label{eq: minimum energy}
 \frac{E_{b}}{N_{0}{_{\mathrm{min}}}}&=\lim_{p_{r} \rightarrow 0}\frac{p_{r}}{R_{lk}\left( p_{r}, \alpha \right)}=\frac{1}{\dot{R}_{lk}\left( 0, \alpha \right)},\\
 S_{0}&=-\frac{2\left[ \dot{R}_{lk}\left( 0, \alpha \right)\right]^{2}}{\ddot{R}_{lk}\left( 0, \alpha \right)}\mathrm{ln}2.\label{eq: wideband_slope}
\end{align}

It is worthwhile to mention that the 
wideband slope $S_{0}$ enables us to study  the  growth of the spectral efficiency with $\frac{E_{b}}{N_{0}}$ in the wideband regime. Basically, $S_{0}$ represents the increase of bits per
second per hertz per $3$ dB of ${E_{b}}$ achieved at $\frac{E_{b}}{N_{0}{_{\mathrm{min}}}}$, and since it is  invariant to channel gain, it is not necessary to distinguish between transmitted and received $S_{0}$.

\begin{theorem}\label{Theo 3}
In the low-SNR regime, the achievable uplink rate from the $k$th
terminal in the $l$th cell to its BS, under delayed
channels, can be represented by the minimum transmit energy per
information bit, $\frac{E_{b}}{N_{0}{_{\mathrm{min}}}}$, and the
wideband slope $S_{0}$, respectively, given by
 \begin{align}
  &\frac{E_{b}}{N_{0}{_{\mathrm{min}}}}=\frac{\mathrm{ln}2}{\alpha^2\left( N-K+1 \right)\hat{\beta}_{llk}}\\
&S_{0}\!= \!\frac{-2\!\left( N\!-\!K\!+\!1 \right)/\left(
N\!-\!K\!+\!2 \right)}{\alpha^{4} \!+\!2 \alpha^{2} C \!\left(\!
N\!-\!K\!+\!3 \!\right)\!+\!\frac{2}{N-K+2}
\!\sum\limits_{p=1}^{\varrho\left(\!
\mathbf{\mathcal{A}}_k\!\right)}
      \!  \sum\limits_{q=1}^{\tau_p \left(\! \mathbf{\mathcal{A}}_k\!\right)} \!
           \! \frac{
                \mathcal{X}_{p,q}\! \left(\! \mathbf{\mathcal{A}}_k\!\right)\mu_{k,p}^{-q}q
            }{
                \hat{ \beta}_{llk}\left(
                    q-1
                \right)
                !
            }}.
            \end{align}
\begin{proof}
See Appendix~\ref{sec:low snr}.
\end{proof}
\end{theorem}

Interestingly, both the minimum transmit energy per information
bit and the wideband slope depend on channel aging by means of
$\alpha$. In particular, as $\alpha$ decreases, both metrics
increase.

\subsection{Large Antenna Limit Analysis}
We next investigate asymptotic performance when $N$ and/or $K$
grow large: i) the number of BS antennas $N$ goes
infinity, while $K$ is fixed, and ii) both the number of terminals
$K$ and the number of BS antennas $N$ grow large, but
their ratio $\kappa=\frac{N}{K}$ is kept fixed. 
Furthermore, the  power scaling law is also studied.

\subsubsection{$N \to \infty$ with fixed $p_{r}$ and $K$}

Note that an Erlang distributed RV, $X_k[n-1]$, with shape
parameter $N-K+1$ and scale parameter $\hat{\beta}_{llk}$ can be
expressed as a sum of independent normal RVs $W_1[n-1], W_2[n-1],
..., W_{2\left(N-K+1\right)}[n-1]$ as follows:
\begin{align}\label{eq large N 1}
    X_k[n-1]
    =
        \frac{\hat{\beta}_{llk}}{2}
        \sum_{i=1}^{2\left(N-K+1\right)}
        W_i^2[n-1].
\end{align}

Substituting \eqref{eq large N 1} into \eqref{eq Prop1 1}, and
using the law of large numbers, the nominator and the first term
of the denominator in \eqref{eq Prop1 1} converge almost surely to
$\alpha^2p_{r}\hat{\beta}_{llk}/2$ respectively $\alpha^2p_{r} C
\hat{\beta}_{llk}/2$ as $N \to \infty$, while the second term of
the denominator goes to $0$. As a result, we have
\begin{align} \label{eq large N 2f}
\gamma_k
     \mathop \rightarrow \limits^{\tt a.s.}
           \frac{1}{C}, \quad \text{as
          $N\to\infty$},
\end{align}
where  $\mathop \rightarrow \limits^{\tt a.s.}$ denotes
almost sure convergence \cite{Vaart,Billingsley}. The bounded SINR is expected because it
is well known that, as $N\to\infty$, the intra-cell interference
and noise disappear, but the inter-cell interference coming from pilot
contamination remains.

\subsubsection{$ K, N \to \infty$ with fixed $p_{r}$ and $\kappa={N}/{K}$}
In practice, if the number of served terminals $K$ in each cell of
next generation systems is  not much less than the number of base
station antennas $N$, then the application of the law of
numbers does not hold because the channel vectors between the
BS and the terminals are not anymore pairwisely
orthogonal. This, in turn, induces new properties in the scenario
under study, which are going to be revealed after the following
analysis. Basically, we will derive the deterministic
approximation $\bar{\gamma}_k$ of the SINR ${\gamma}_k$ such that
\begin{align}
{\gamma}_k- \bar{\gamma}_k\xrightarrow[ N \rightarrow
\infty]{\mbox{a.s.}} 0.
\end{align}

\begin{theorem}
The deterministic equivalent $\bar{\gamma}_k$ of the uplink SINR
between the $k$th terminal in the $l$th cell and its BS
is given by
\begin{align}\label{eq large k theorem}
\bar{\gamma}_k= \frac{
        \alpha^2\hat{\beta}_{llk}\left(\kappa-1 \right)
        }{
        \alpha^2C\hat{\beta}_{llk}\left(\kappa-1 \right)
+
  \sum_{i=1}^{L} \frac{1}{K}{\tt Tr} \tilde{\B{D}}_{li}
        }.
\end{align}
\begin{proof}
Since $\hat{\B{Y}}_i[n]\sim \CG{\B{0}}{\tilde{\B{D}}_{li}}$, it
can be rewritten as:
\begin{align}\label{eq Yi}
\hat{\B{Y}}_i[n]&=  \B{a}_{i}^\H \tilde{\B{D}}_{li}^{\frac{1}{2}},
\end{align}
where $\B a_{i}\!\!~\!\sim \CG{\B{0}}{\Id_{K}}$. By
substituting~\eqref{eq large N 1} and \eqref{eq Yi} into~\eqref{eq
Prop1 1}, we have
\begin{align} \label{eq large k}
    \gamma_k
    &=
        \frac{
            \alpha^2p_{r}
            \frac{\hat{\beta}_{llk}}{2}
            \sum_{i=1}^{2\left(N-K+1\right)}
            W_i^2[n-1]
            }{
           \alpha^2p_{r} C
            \frac{\hat{\beta}_{llk}}{2}\!\!
            \sum\limits_{i=1}^{2\left(N-K+1\right)}\!\!
            W_i^2[n-1]
            +p_r
            \sum\limits_{i =1}^{L}
                    \B{a}_i^H
                    \tilde{\B{D}}_{li}
                    \B{a}_i
        +
            1
            }.
\end{align}
Next, if we divide both the nominator and denominator of \eqref{eq
large k} by $2\left( N-K+1 \right)$ and by using
\cite[Lemma~1]{Truong}, under the assumption that
$\tilde{\B{D}}_{li}$ has uniformly bounded spectral norm with
respect to $K$, we arrive at the desired result \eqref{eq large k
theorem}.
\end{proof}
\end{theorem}

\begin{remark}
Interestingly, in contrast to~\eqref{eq large N 2f}, the SINR is
now affected by intra-cell interference  as well as inter-cell
interference and it does not depend on the transmit power. In fact,
the former justifies the latter, since both the desired and
interference signals are changed by the same factor, if each
terminal changes its power. Note that the interference terms
remain because they depend on both $N$ and $K$; however, the
dependence of thermal noise only from $N$ makes it vanish. As
expected,~\eqref{eq large k theorem} coincides with~\eqref{eq
large N 2}, if $N\gg K$, i.e., when $\kappa \to \infty$, the SINR
goes asymptotically to ${1}/{C}$. The result \eqref{eq large k theorem} can be applied for finite
$N$ and $K$  by adjusting parameter $\kappa$~\cite{Truong,Papazafeiropoulos2,Papazafeiropoulos1,Papazafeiropoulos3
}.
\end{remark}

Next, the deterministic equivalent rate can be obtained by means
of the dominated convergence~\cite{Billingsley} and the continuous
mapping theorem~\cite{Vaart} as
\begin{align}
 R_{lk}(p_r,\alpha)- \log_2\left( 1+ \bar{\gamma}_k\right) \xrightarrow[ N \rightarrow \infty]{\mbox{a.s.}} 0.
\end{align}

\subsubsection{Power-Scaling Law}

Let  $p_{r} = E/\sqrt{N}$, where $E$ is fixed regardless
of $N$. Given that  $\hat{\beta}_{llk}$ depends on
$p_{\mathrm{tr}}=\frac{\tau E}{\sqrt{N}}$, we have that for fixed $K$ and $N\to\infty$,
\begin{align} \label{eq large N 2}
\gamma_k \mathop \rightarrow \limits^{\tt a.s.}
\frac{\alpha^{2}\tau E^2 {\beta}_{llk}^2}{\alpha^{2} \tau E^2 C
{\beta}^2_{llk}+1},
\end{align}
which is a non-zero constant. This implies that, we can reduce the
transmit power proportionally to $1/\sqrt{N}$, while retaining a
given quality-of-service. In the case where the BS has
perfect CSI and where there is no relative movement of the
terminals, the result \eqref{eq large N 2} is identical with the
result in \cite{Mathaiou:ZF_receivers}.

\begin{figure}[t]
    \centering
    \centerline{\includegraphics[width=0.48\textwidth]{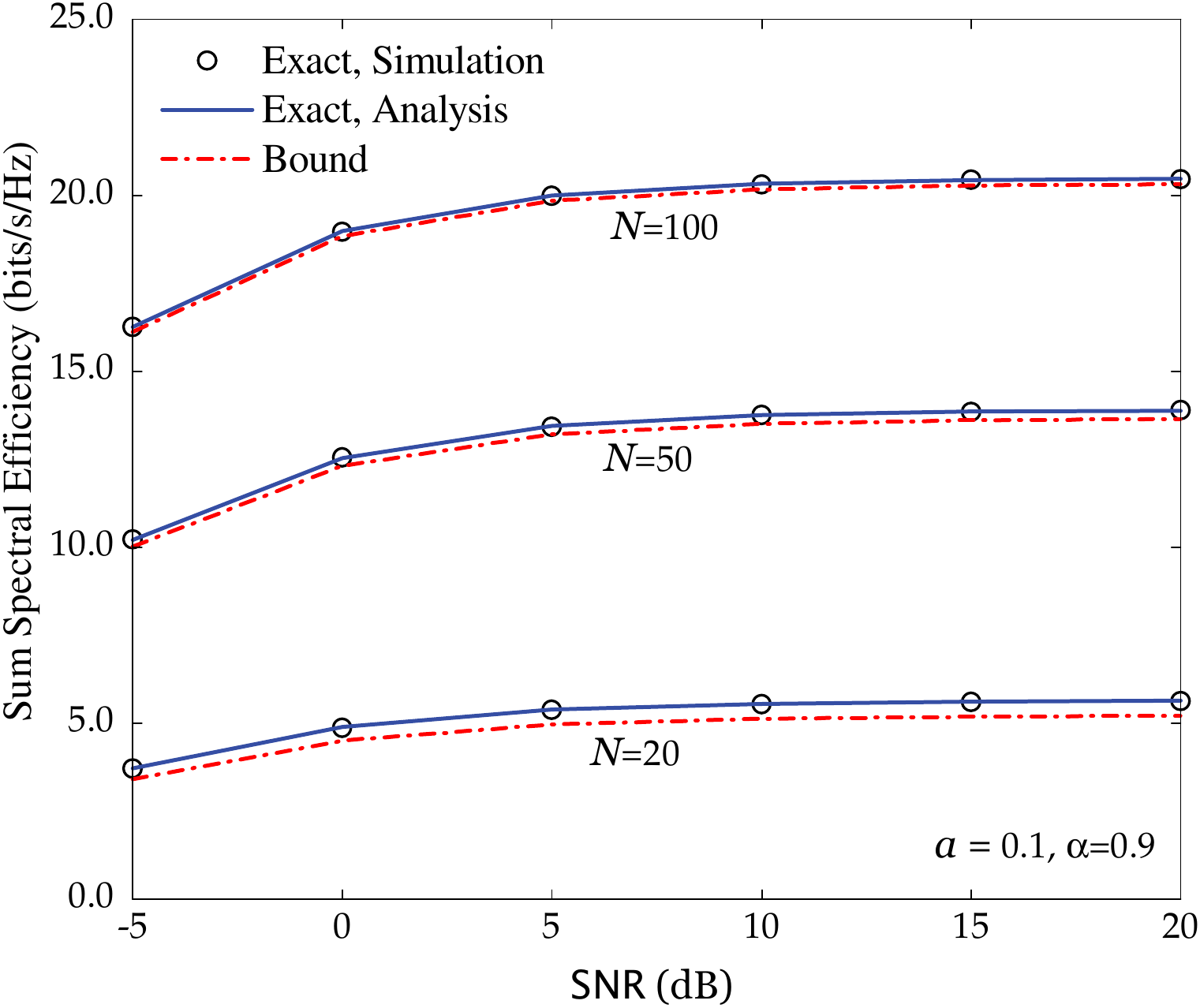}}
    \caption{Sum spectral efficiency  versus $\mathsf{SNR}$ for different $N(a=0.1$ and $\alpha=0.9)$.}
    \label{fig:1}
\end{figure}

\begin{figure}[t]
    \centering
    \centerline{\includegraphics[width=0.48\textwidth]{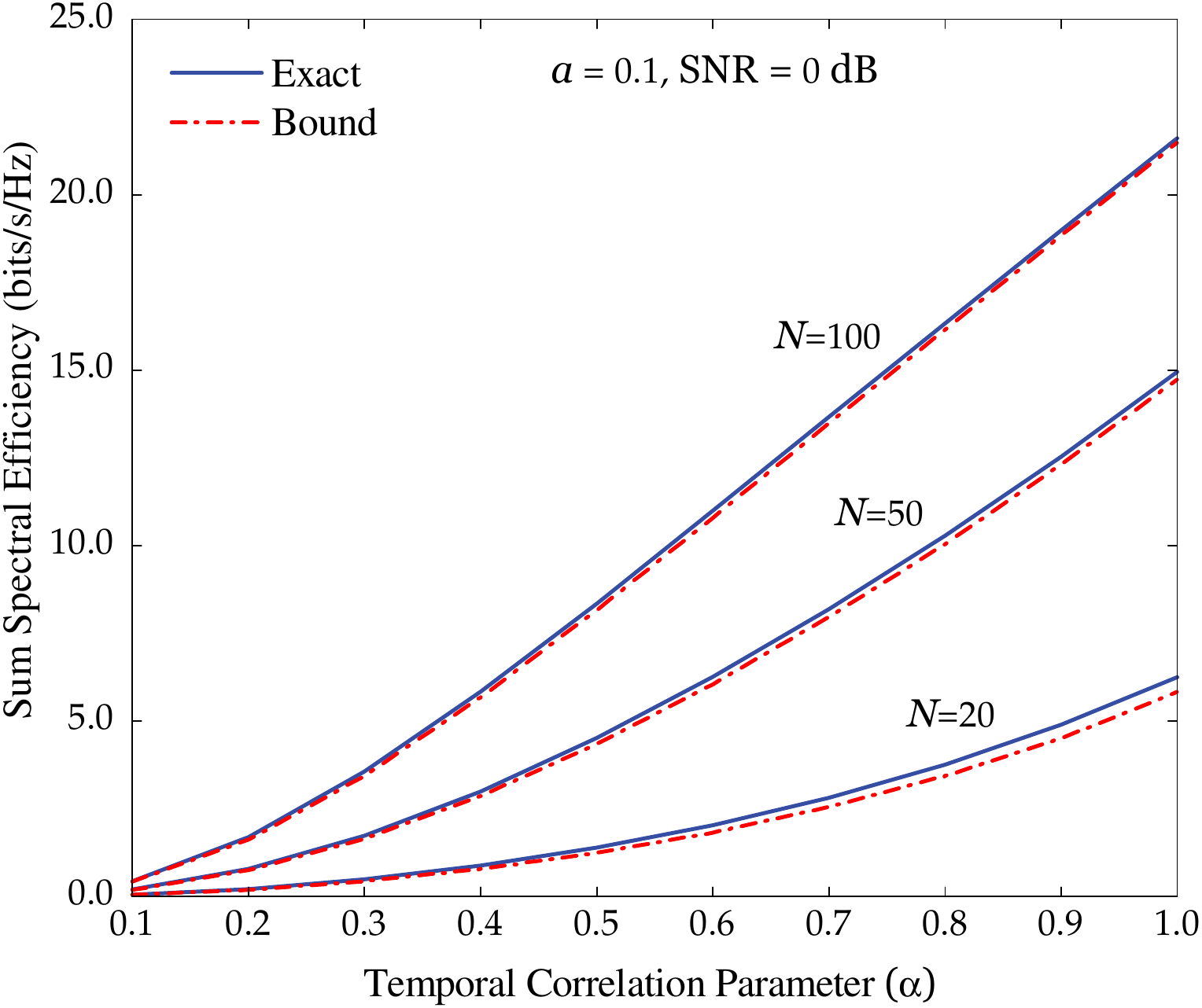}}
    \caption{Sum spectral efficiency  versus $\alpha$ for different $N(a=0.1$ and $\mathsf{SNR}=0\,$dB).}
    \label{fig:2}
\end{figure}

\begin{figure}[t]
    \centering
    \centerline{\includegraphics[width=0.48\textwidth]{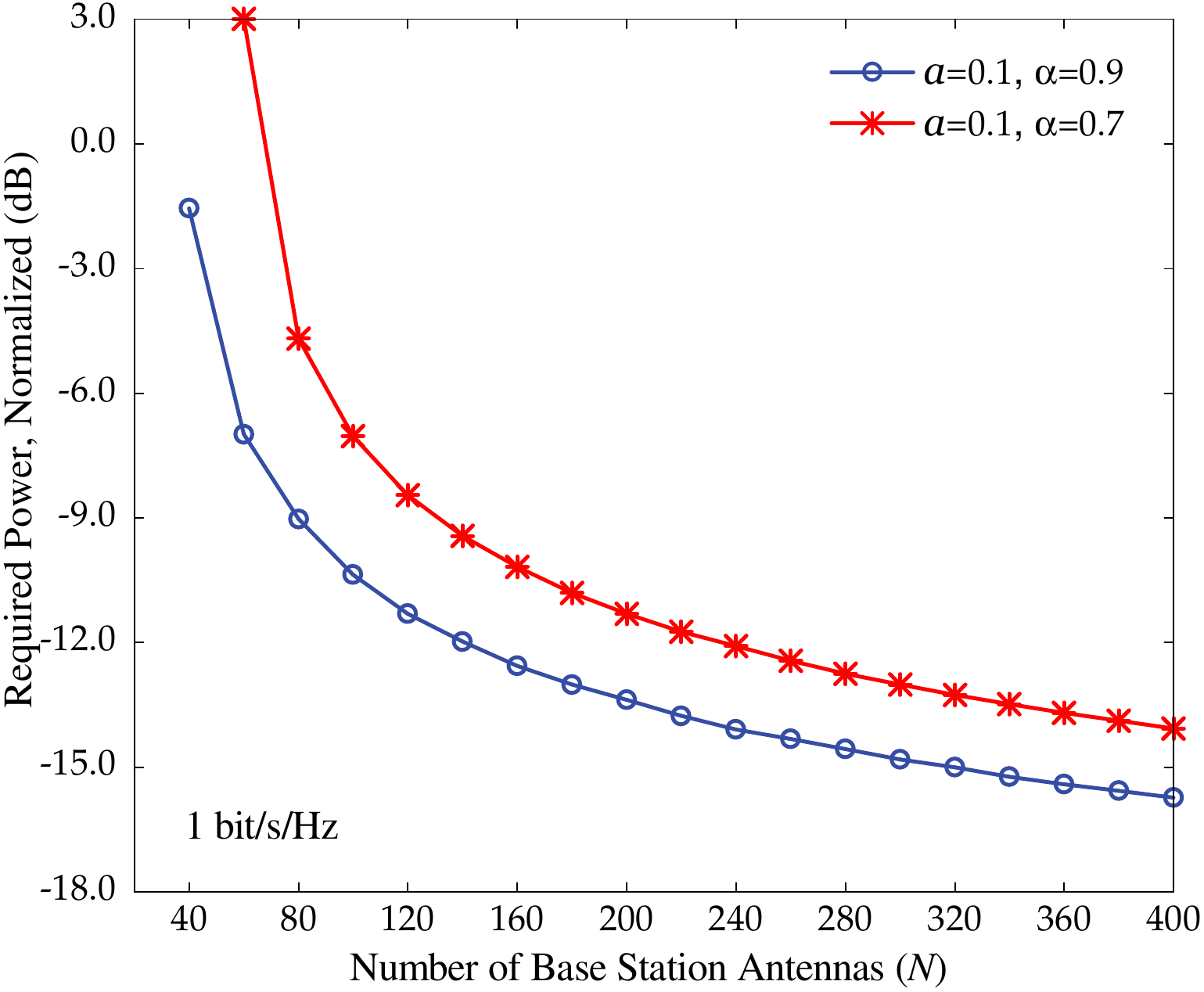}}
    \caption{Transmit power required to achieve 1 bit/s/Hz per terminal versus $N(a=0.1$, $\alpha=0.7$ and $\alpha=0.9)$.}
    \label{fig:3}
\end{figure}

\begin{figure}[t]
    \centering
    \centerline{\includegraphics[width=0.48\textwidth]{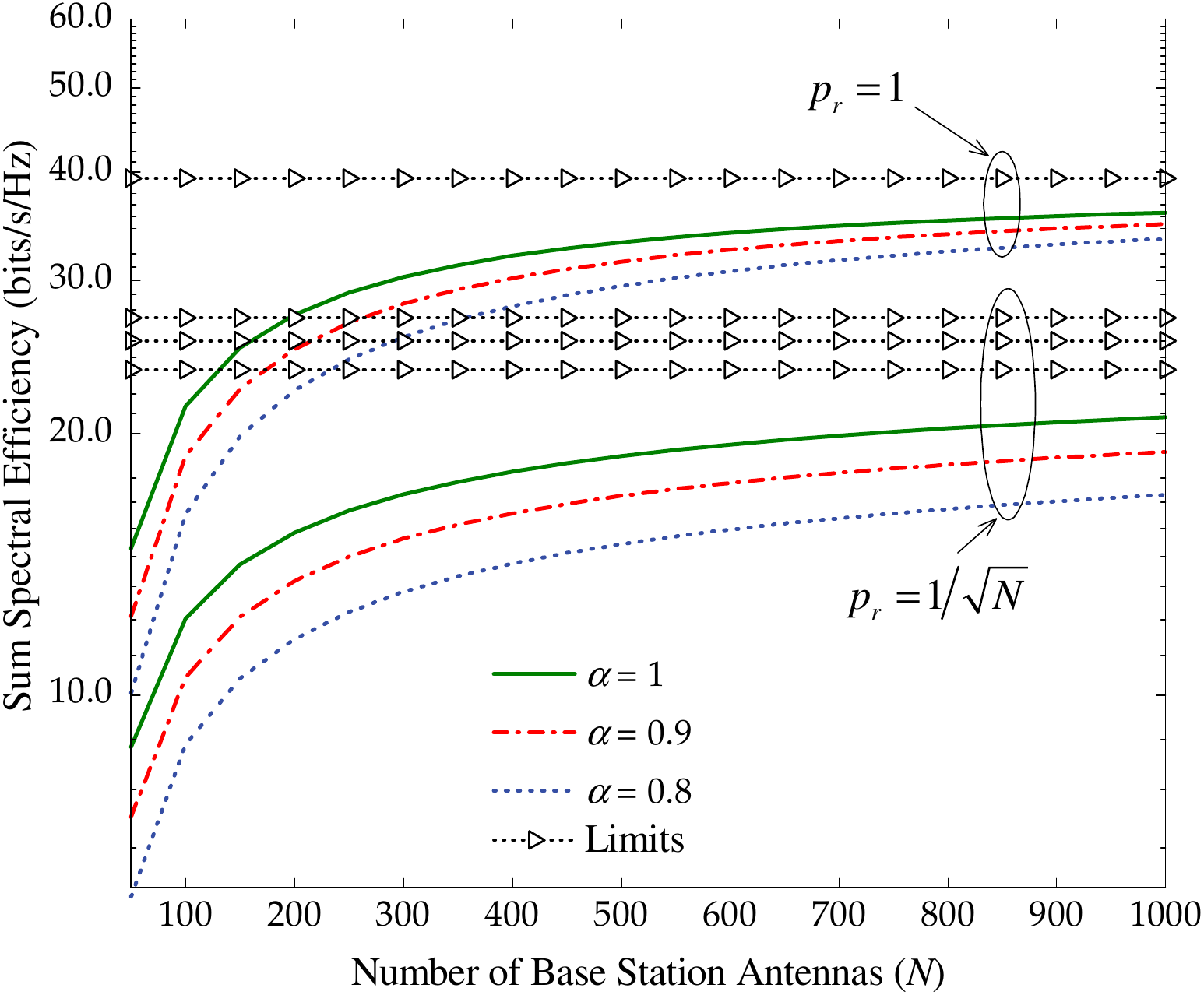}}
    \caption{Sum spectral efficiency  versus $N$ for different $\alpha$.}
    \label{fig:4}
\end{figure}

\begin{figure}[t]
    \centering
    \centerline{\includegraphics[width=0.465\textwidth]{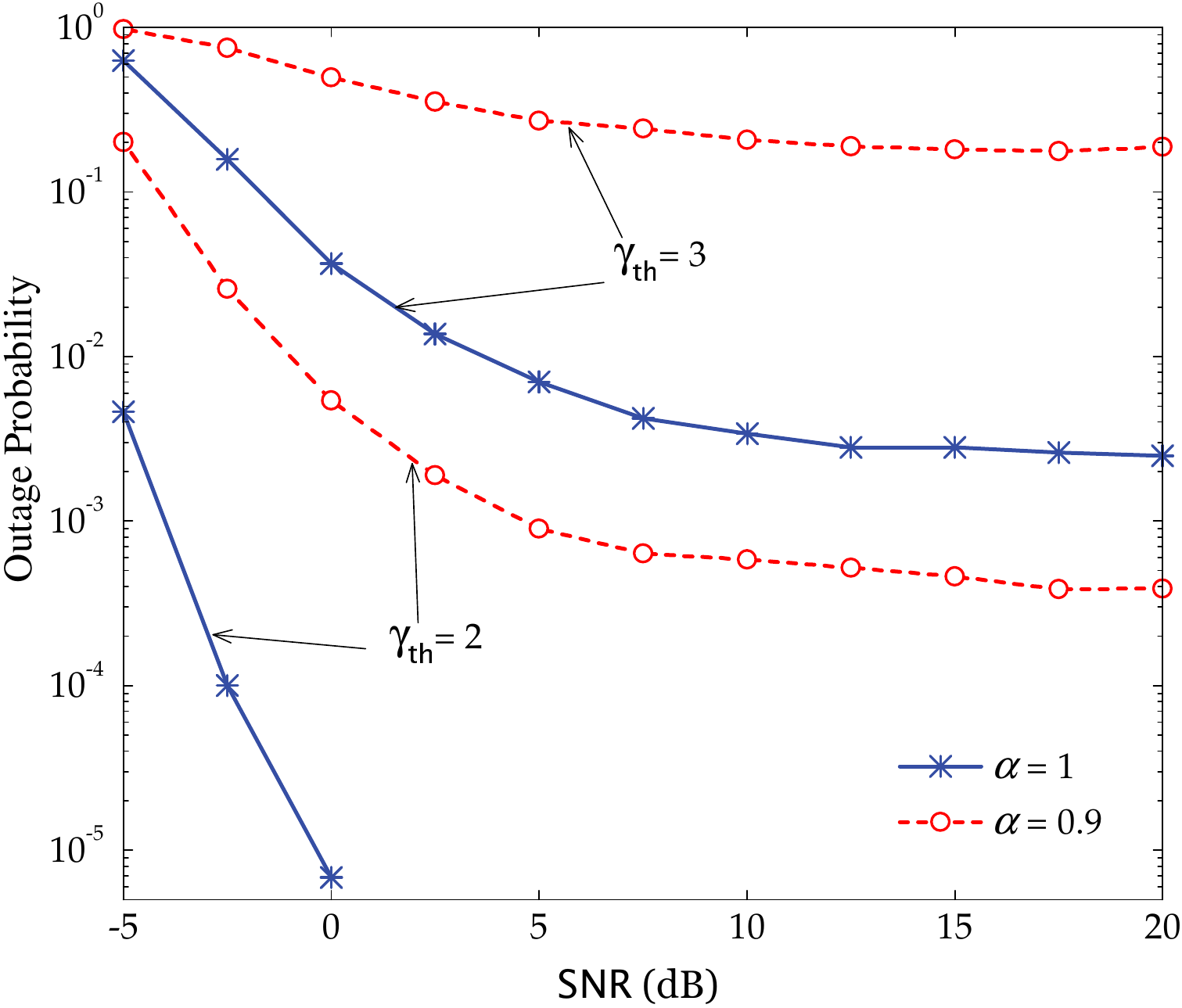}}
    \caption{Outage probability  versus $\mathsf{SNR}$ for different $\alpha$ and $\gamma_{\tt th}$ ($N=100$).}
    \label{fig:5}
\end{figure}

\section{Numerical Results}\label{Numerical} 
In this section, we provide numerical results to corroborate our
analysis. We deploy a cellular network having  $L=7$ cells, each
cell has $K=10$ terminals. We choose the frame length is
$T=200$ symbols. For each frame, a
duration of length $\tau=K$ symbols is used for uplink training.
Regarding the large-scale fading coefficients $\beta_{lik}$, we
employ a simple model: $\beta_{llk}=1$ and $\beta_{lik}=a$, for
$k=1, \ldots, K$, and $i\neq l$. For this simple model, $a$ is
considered as an inter-cell interference factor. In all examples,
we choose $a=0.1$. Furthermore, we define $\mathsf{SNR}\triangleq
p_{r}$.

In the following, we scrutinize  the sum-spectral efficiency, defined
as:
\begin{align} \label{eq num 1}
\mathcal{S}_l \triangleq \left(1-\frac{\tau}{T}\right)\sum_{k=1}^K
 R_{lk}\left(p_r,\alpha \right),
\end{align}
where $ R_{lk}\left(p_r,\alpha \right)$ is given in \eqref{Rate
prop}.

Figure~\ref{fig:1} represents the sum spectral efficiency as a
function of SNR for different $N$, with the intercell interference
factor $a=0.1$ and the temporal correlation parameter
$\alpha=0.9$. The ``Exact, Simulation'' curves are generated via
\eqref{eq: SINR} using Monte-Carlo simulations, the ``Exact,
Analysis'' curves are obtained by using \eqref{Rate prop}, while
the ``Bound'' curves are derived by using the bound formula given
in Proposition~\ref{Prop BoundRate 1}. The exact agreement between the
 simulated and analytical results validates our analysis and shows that the proposed bound is very tight,
especially for a large number of BS antennas. Furthermore,
as in the analysis, at high SNR, the sum spectral efficiency
saturates. To enhance the system performance, we can add more
antennas at the BS. At $\mathsf{SNR}=5$dB, if we
increase $N$ from $20$ to $50$ or from $20$ to $100$, then the
sum spectral efficiency can be increased by the factors of $2.5$
or $5.5$.

Next, we study the effect of the temporal correlation
parameter $\alpha$ on the system performance and examine the
tightness of our proposed bound  in Proposition~\ref{Prop
BoundRate 1}. Figure~\ref{fig:2} shows the sum spectral efficiency
versus $\alpha$, for $N=20, 50$, and $100$. Here, we choose
$\mathsf{SNR}=0$dB. When the temporal correlation parameter
decreases (or the time variation of the channel increases), the
system performance deteriorates significantly. When $\alpha$ decreases 
from $1$ to $0.6$, the spectral efficiency is reduced by a factor
of $2$. In addition, at low $\alpha$, using more antennas at the
BS does not help much in the improvement of the system
performance. Regarding the tightness of the proposed bound, we can
see that the bound is very tight across the entire temporal
correlation range.

Figure~\ref{fig:3} depicts the transmit power, $p_{r}$, that is
required to obtain $1$ bit/s/Hz per terminal, for $\alpha=0.7$ and
$0.9$. As expected, the required transmit power reduces
significantly when the number of BS antennas increases.
By doubling the number of BS antennas, we can cut back
$p_{r}$ by approximately $1.5$~dB. This property is identical
to the results of~\cite{Ngo_Energy}.

To further verify our analysis on large antenna limits, we
consider Figure~\ref{fig:4}. Figure~\ref{fig:4} shows the sum
spectral efficiency versus the number of BS antennas for
different values of $\alpha$, and for two cases: the transmit
power, $p_{r}$, is fixed regardless of $N$, and the transmit power
is scaled as $p_{r}=1/\sqrt{N}$. The ``Limits" curves are derived
via the results obtained in Section~III-C. As expected, as the
number of the BS antennas increases, the sum spectral
efficiencies converge to their limits. When the transmit power is
fixed, the asymptotic performance (as $N\to\infty$) does not
depend on the temporal correlation parameter. By contrast, when
the transmit power is scaled as $1/\sqrt{N}$, the asymptotic
performance depends on $\alpha$.

Finally, we shed light on the outage performance versus SNR at $N=100$,
for different temporal correlation parameters ($\alpha=1$, and
$0.9$), and for different threshold values ($\gamma_{\tt th}=2$,
and $3$). See Figure~\ref{fig:5}. We can observe that the outage
probability strongly depends on $\alpha$. At $\mathsf{SNR}=0$ dB,
by reducing $\alpha$ from 1 to $0.9$, the outage probability
increases from $7\times 10^{-6}$ to $5\times 10^{-3}$, and from
$3\times 10^{-2}$ to $5\times 10^{-1}$ for $\gamma_{\tt th}=2$,
and $3$, respectively. In addition, the outage probability
significantly improves when the threshold values are slightly
reduced. The reason is that, with large antenna arrays,
the channel hardening occurs, and hence, the SINR concentrates
around its mean. As a result, by slightly reducing the threshold
values, we can obtain a very low outage probability.

\section{Conclusions}\label{Conclusions} 
This paper analyzed the uplink performance of cellular networks
with zero-forcing receivers, coping with the well-known
pilot contamination effect and the unavoidable, but less studied,
channel aging effect. The latter effect, inherent in the vast
majority of practical propagation environments, stems from the terminal
mobility. Summarizing the main contributions of this work,  new
analytical closed-form expressions for the PDF of the SINR and the
corresponding achievable 
ergodic rate that hold for any finite
number of BS antennas were derived. Moreover, a
complete investigation of the low-SNR regime took place.
Nevertheless, asymptotic expressions in the large numbers of
antennas/terminals limit were also obtained, as well as the
power-scaling law was studied. As a final point, numerical
illustrations represented how the channel aging phenomenon affects
the system performance for a finite and an infinite number of antennas.
Notably, the outcome is that large number of antennas should be
preferred even in time-varying conditions.

\appendix

\subsection{Proof of Proposition~\ref{Prop 1}} \label{appproofPro1}

By dividing the numerator and denominator of \eqref{eq: SINR} by
$\left\| \left[\hat{\B{G}}_{ll}^{\dagger}[n-1]\right]_k
\right\|^2$, we have
\begin{align} \label{eq Rate 1}
 \!\!\!\gamma_k
   =
        \frac{
            \alpha^2\Pu
            \left\| \left[\hat{\B{G}}_{ll}^{\dagger}[n-1]\right]_k
          \right\|^{-2}
            }{
           \alpha^2 \Pu C
            \left\|\! \left[\hat{\B{G}}_{ll}^{\dagger}[n\!-\!1]\right]_k
            \right\|^{-2}
            \!\!+\!
        \Pu \sum_{i =1}^{L}\!
                \left\|
                    \hat{\B{Y}}_i[n]
                \right\|^2
\!+\!
            1
            },
\end{align}
where
\begin{align} \label{eq Rate 1aa}
C
    &\triangleq
    \sum_{i \ne
l}^{L}\!\left\|\!\left[\hat{\B{G}}_{ll}^{\dagger}[n-1]\right]_k
\!\hat{\B{G}}_{li}[n-1] \right\|\!=\! \sum_{i \ne
l}^{L}\!\left(\!\frac{\beta_{lik}}{\beta_{llk}}\! \right)^2,\\
\hat{\B{Y}}_i[n]
    &\triangleq
    \frac{\left[\hat{\B{G}}_{ll}^{\dagger}[n-1]\right]_k
\tilde{\B{E}}_{li}[n]}{\left\|\left[\hat{\B{G}}_{ll}^{\dagger}[n-1]\right]_k
\right\|}. \label{eq Rate 1bb}
\end{align}
Note that the last equality in \eqref{eq Rate 1aa} follows
\eqref{interchannelEstimated}. Since
$$\left\|
\left[\hat{\B{G}}_{ll}^{\dagger}[n-1]\right]_k \right\|^2 =
\left[\left(\hat{\B{G}}_{ll}^\H[n-1] \hat{\B{G}}_{ll}
[n-1]\right)^{-1} \right]_{kk},$$ $\left\|
\left[\B{G}_{ll}^{\dagger}[n-1]\right]_k \right\|^{-2}$  has an
Erlang distribution with shape parameter $N\!-\!K\!+\!1$ and scale
parameter $\hat{\beta}_{llk}$ \cite{GHP:02:CL}.
\footnote{
The Erlang and Gamma distributions, having the same parameters, coincide, if the shape parameter is an integer. More concretely, if $N\!-\!K\!+\!1$ is an integer: $X \!\sim \!  \Gamma(N\!-\!K\!+\!1, \hat{\beta}_{llk} ) $ (gamma distribution), then  $X\! \sim\!  \mathrm{Erlang}(N\!\!-K\!+\!1,\hat{\beta}_{llk})$.} Therefore,
\begin{align} \label{eq PDF1 1a}
    \left\| \left[\hat{\B{G}}_{ll}^{\dagger}[n-1]\right]_k \right\|^{-2}
    \mathop \sim \limits^{\tt d}
    X_k[n-1].
\end{align}
Furthermore, for a given
$\left[\hat{\B{G}}_{ll}^\dagger[n-1]\right]_k$, $\hat{\B{Y}}_i[n]$
is a complex Gaussian vector with a zero-mean and covariance
matrix $\tilde{\B D} _{li}$ which is independent of
$\left[\hat{\B{G}}_{ll}^\dagger[n-1]\right]_k$. Thus,
$\hat{\B{Y}}_i[n]\sim \CG{\B{0}}{\tilde{\B{D}}_{li}}$, and is
independent of $\left[\hat{\B{G}}_{ll}^\dagger[n-1]\right]_k$. As
a result, $\sum_{i=1}^{L}\left\| \hat{\B{Y}}_i[n] \right\|^2$ is
the sum of $K L$ independent but not necessarily identically
distributed exponential RVs. From \cite[Theorem~2]{BSW:07:WCOM},
we have that
\begin{align}
 \sum_{i= l}^{L}
                \left\|
                    \hat{\B{Y}}_i[n]
                \right\|^2
   \overset{\tt d}{\sim}
    Y_k[n].\label{eq PDF1 y}
\end{align}
Combining \eqref{eq Rate 1}--\eqref{eq PDF1 y}, we arrive at
\eqref{eq Prop1 1} in Proposition~\ref{Prop 1}.

\subsection{Proof of Theorem~\ref{Theo 1}} \label{sec:proof:prop
rate1} The achievable uplink ergodic rate of the $k$th terminal in
the $l$th cell is given by
\begin{align}
     &R_{lk}\!\left(\! p_{r},\alpha\! \right)
    \!=\!
        \EXs{X_k, Y_k\!\!\!}{
            \!\log_2\!\!
            \left(\!\!
                1\!+\!
                \frac{
                    p_{r} \alpha^2 X_k[n-1]
                    }{
                    p_{r} \alpha^2 C X_k[n\!-\!1] \!+\! p_{r} Y_k[n]\!+\! 1
                }
            \!\right)
        } \nonumber
    \\
 &\!=\!
    \int_0^{\infty}\!\!\!\!
    \int_0^{\infty}\!\!\!
           \log_2
            \left(\!
                1\!+\!
                \frac{
                    p_{r} \alpha^2 x
                    }{
                    p_{r} \alpha^2 C x + p_{r} y+ 1
                }
            \!\right)\!
            \PDF{X_k}{x}
            \PDF{Y_k}{y}
    dx dy.
  \nonumber
\end{align}

Using \eqref{eq PDF1 1} and \eqref{eq PDF1 3}, we obtain
\begin{align}
     &R_{lk}\!\left(\! p_{r},\alpha\! \right) =
        \sum_{p=1}^{\varrho\left( \mathbf{\mathcal{A}}_k\right)}
        \sum_{q=1}^{\tau_p \left( \mathbf{\mathcal{A}}_k\right)}\!
            \frac{
                 \mathcal{X}_{p,q} \left( \mathbf{\mathcal{A}}_k\right)
               \mu_{k,p}^{-q} \log_2 e
            }{
                \left(
                    q\!-\!1
                \right)
        !
                \left(N\!-\!K\right)!
                \hat{\beta}_{llk}^{N-K+1}
            }
        \nonumber
       \\
       &  \!\times\!
    \!\int_0^{\infty}\!\!\!
    \!\int_0^{\infty}\!\!
            \ln\!\!
            \left(\!\!
                1\!+\!
                \frac{
                    p_{r} \alpha^2 x
                    }{
                    p_{r} \alpha^2 C x \!+\! p_{r} y \!+ \!1
                }
            \!\right)\!
            x^{N\!-\!K} e^{\frac{-x}{\hat{\beta}_{llk}}}
            y^{q-1}
           e^{\frac{-y}{\mu_{k,p}}}      dx dy\nn\\
           &= \sum_{p=1}^{\varrho\left( \mathbf{\mathcal{A}}_k\right)}
        \sum_{q=1}^{\tau_p \left( \mathbf{\mathcal{A}}_k\right)}\!
            \frac{
                 \mathcal{X}_{p,q} \left( \mathbf{\mathcal{A}}_k\right)
               \mu_{k,p}^{-q} \log_2 e
            }{
                \left(
                    q\!-\!1
                \right)
        !
                \left(N\!-\!K\right)!
                \hat{\beta}_{llk}^{N-K+1}
            }
        \nonumber
       \\
       &  \!\times\! \bigg(
    \!\underbrace{\int_0^{\infty}\!\!\!\!
    \!\int_0^{\infty}\!
            \ln\!\!
            \left(\!
                1\!+\!
                \frac{
                    p_{r} \alpha^2 \left(\! C\!+\!1 \!\right)x
                    }{
                     p_{r} y \!+ \!1
                }
            \!\right)\!
            x^{N\!-\!K} e^{\frac{-x}{\hat{\beta}_{llk}}}
            y^{q-1}
           e^{\frac{-y}{\mu_{k,p}}}      dx dy}_{\triangleq \mathcal{I}_1} \nn\\
           &\!-\!
           \!\underbrace{\int_0^{\infty}\!
    \!\int_0^{\infty}\!
            \ln
            \left(\!
                1\!+\!
                \frac{
                    p_{r} \alpha^2 C x
                    }{
                   \ p_{r} y \!+ \!1
                }
            \!\right)\!
            x^{N\!-\!K} e^{\frac{-x}{\hat{\beta}_{llk}}}
            y^{q-1}
           e^{\frac{-y}{\mu_{k,p}}}dx dy}_{\triangleq \mathcal{I}_2}\!\!\bigg)\nonumber \\
           &=
\sum_{p=1}^{\varrho\left( \mathbf{\mathcal{A}}_k\right)}
        \sum_{q=1}^{\tau_p \left( \mathbf{\mathcal{A}}_k\right)}\!
            \frac{
                 \mathcal{X}_{p,q} \left( \mathbf{\mathcal{A}}_k\right)
               \mu_{k,p}^{-q} \log_2 e
            }{
                \left(
                    q\!-\!1
                \right)
        !
                \left(N\!-\!K\right)!
                \hat{\beta}_{llk}^{N-K+1}
            }    \left(  \mathcal{I}_{1}-\mathcal{I}_{2} \right).\label{Rate 1}
\end{align}

We first derive $\mathcal{I}_{1}$ by evaluating the integral over
$x$. By using \cite[Eq.~(4.337.5)]{GR:07:Book}, we obtain
\begin{align} \label{Rate 2}
  \mathcal{I}_{1}
    &=   \sum_{t=0}^{N-K} \!\int_0^{\infty}\!
        \left[
-f(y)^{N-K-t}
                e^{-f(y)}
                \mathrm{Ei}\left(f(y) \right)
        \right.
        \nonumber
        \\
        &
            \left.
                +
                \sum_{u=1}^{N-K-t}
                \left(u-1 \right)!
                f(y)^{N-K-t-u}
            \right]
           y^{q-1}
           e^{\frac{-y}{\mu_{k,p}}}
    dy,
\end{align}
where $f(y)\triangleq -\frac{ p_{r}  y +1}{\hat{\beta}_{llk} p_{r}
\alpha^2  \left( C+1 \right)}$. Using \cite[Lemma~1]{Duong} and
\cite[Eq.~(39)]{KA:06:WCOM}, we can easily obtain
$\mathcal{I}_{1}$ as given in \eqref{eq Prop2 1a}. Similarly, we
obtain $\mathcal{I}_{2}$ as given in \eqref{eq Prop2 1b}.
Substitution of  $\mathcal{I}_{1}$ and $\mathcal{I}_{2}$
into~\eqref{Rate 1} concludes the proof.

\subsection{Proof of Proposition~\ref{Prop BoundRate
1}}\label{propBound Rate}

By using  Jensen's inequality, we have
\begin{align} \label{Rate 1b}
      R_{lk}\left( p_{r},\alpha \right)
&=\EX{\log_2\left(1+ \gamma_{k}\right)}
=\EX{\log_2\left(1+ \frac{1}{1/\gamma_{k}}\right)}\nn\\
&\ge \log_2\left(1+ \frac{1}{\EX{1/\gamma_{k}}}\right) \triangleq
R_{L}\left( p_{r},\alpha \right).
\end{align}
To compute $R_{L}\left( p_{r},\alpha \right)$, we need to compute
$\EXs{}{1/\gamma_{k}}$. From \eqref{eq Rate 1}, we have
\begin{align}
 &\EXs{}{\frac{1}{\gamma_{k}}}=C+\frac{1}{\alpha^{2}}\sum_{i=1}^{L}\EXs{}{\bigg\|\left[\hat{\B{G}}_{ll}^{\dagger}[n-1]\right]_k \tilde{\B{E}}_{li}[n]\bigg\|^{2}}\nn\\
 &\hspace{3.5cm}+\frac{1}{\alpha^{2}p_{r}}\EXs{}{\bigg\|\left[\hat{\B{G}}_{ll}^{\dagger}[n-1]\right]_k \bigg\|^{2}}\nn\\
  &\!=\!C\!+\!\frac{1}{\alpha^{2}}\!\! \EXs{}{\bigg\|\!\!\left[\!\hat{\B{G}}_{ll}^{\dagger}[n\!-\!1]\!\right]_k \!\bigg\|^{2}}\!\!\left(\sum_{i=1}^{L}\!\sum_{k=1}^{K}\! \left(\! \beta_{lik}\!-\!\alpha^{2}\hat{\beta}_{lik} \!\right)\!+\!\frac{1}{p_{r}} \!\right)\nn\\
  &\!=\!C\!+\!\frac{1}{\left(\! N\!-\!K \!\right)\! \alpha^{2}\hat{ \beta}_{llk}}\left(\sum_{i=1}^{L}\sum_{k=1}^{K}\! \left( \beta_{lik}\!-\!\alpha^{2}\hat{\beta}_{lik} \!\right)\!+\!\frac{1}{p_{r}} \right).\label{Rate 1d}
 \end{align}
 In the third equality of \eqref{Rate 1d}, we have considered the independence between the two variables,
while in the last equality, we have used the following result:
 \begin{align}
  &\EXs{}{\bigg\|\left[\hat{\B{G}}_{ll}^{\dagger}[n-1]\right]_k \bigg\|^{2}}=\EXs{X_{k}}{\frac{1}{X_{k}[n-1]}}\nn\\
  &\!=\! \int_{0}^{\infty}\!\!\!\frac{
            e^{-x/\hat{ \beta}_{llk}}
            }{
            \left(N-K\right)!
           \hat{ \beta}_{llk}^{2}
            }\!\!
        \left(
            \frac{
                x
                }{
               \hat{ \beta}_{llk}
                }
        \right)^{N\!-\!K\!-\!1}\!\!\!\mathrm{d}x\nn\\
        &= \frac{1}{\left( N-K \right) \hat{ \beta}_{llk}}.\label{Rate 1e}
 \end{align}
Note that we have used~\cite[Eq.~(3.326.2)]{GR:07:Book} to obtain
\eqref{Rate 1e}. Thus, the desired result \eqref{eq: LB} is
obtained from~\eqref{Rate 1b} and ~\eqref{Rate 1d}.

\subsection{Proof of Theorem~\ref{theo 2}}  \label{proof:Prop out}

Clearly, from \eqref{eq Prop1 1}, $\gamma_{k} < 1/C$. Thus, if
$\gamma_{{\tt th}} \geq 1/C$, then  $P_{{\tt out}}\left(
\gamma_{{\tt th}}\right) =1$. Hence, we focus on the case where
$\gamma_{{\tt th}} < 1/C$. Taking the probability of the
instantaneous SINR $\gamma_{k}$, given by~\eqref{eq Prop1 1}, we
can determine the outage probability as
\begin{align}
 \!\!\!\!\!\!&P_{{\tt out}}
 \!= \!{\tt Pr}\left( \frac{\alpha^2 p_{r} X_k}{\alpha^2 p_{r} C X_k + p_{r} Y_k+ 1}\leq\gamma_{{\tt
 th}}\right)\nn\\
 &\!=\! \int_0^\infty {\tt Pr}\left( X_k< \frac{\gamma_{{\tt
th}}\left( p_{r} Y_k +1 \right)}{ \alpha^{2}p_{r} - \gamma_{{\tt
 th}} \alpha^{2} p_{r} C}\left.\right|Y_k \right) p_{Y_k}(y)\mathrm{d}y\nn\\
  &\!=\!1\!-\!e^{\frac{-\gamma_{{\tt th}}}{p_{r}\bar{\gamma}_{{\tt th}} }}\!
\sum^{N-K}_{t=0}\!\! \int_{0}^{\infty}\! \!\!e^{
\frac{-y}{\bar{\gamma}_{{\tt th}} }}
 \!\sum^{N-K}_{t=0}\!\!\frac{\left( \frac{ \gamma_{{\tt th}}}{\bar{\gamma}_{{\tt th}}} \right)^{\!\!t}}{t!}\!\left( \!y\!+\!\frac{1}{p_{r}} \!\right)^{\!\!t}\!\!p_{Y_k}(y)\mathrm{d}y \nn\\
 &\!=\!1\!-\!e^{\frac{-\gamma_{{\tt th}}}{p_{r}\bar{\gamma}_{{\tt th}} }}
\sum_{p=1}^{\varrho\left( \mathbf{\mathcal{A}}_k\right)}
\sum_{q=1}^{\tau_p \left( \mathbf{\mathcal{A}}_k\right)}
 \sum_{t=0}^{N-K}
 \mathcal{X}_{p,q} \left(
 \mathbf{\mathcal{A}}_k\right)\frac{\mu_{k,p}^{-q}}{\left(q-1\right)} \frac{\left( \frac{ \gamma_{{\tt th}}}{\bar{\gamma}_{{\tt th}}} \right)^{t}}{t!}\nn\\
 &\hspace{3cm}\times\int_{0}^{\infty} y^{q-1}e^{\frac{-y}{\bar{\gamma}_{{\tt th}})} }
 \left( y+\frac{1}{p_{r}} \right)^{t}\mathrm{d}y \nn\\
 &\!=\!1\!-\!e^{\frac{-\gamma_{{\tt th}}}{p_r\bar{\gamma}_{{\tt th}}}}\!\!
\sum_{p=1}^{\varrho\left( \mathbf{\mathcal{A}}_k\right)}\!
\sum_{q=1}^{\tau_p \left( \mathbf{\mathcal{A}}_k\right)}\!
 \sum_{t=0}^{N-K} \!\!
  \sum_{s=0}^{t}\!\! \binom{t}{s}
 \mathcal{X}_{p,q} \!\left(
 \mathbf{\mathcal{A}}_k\!\right)\!\frac{\Gamma\left( s\!+\!q \right)\! \bar{\gamma}_{{\tt
 th}}^{s+q}}{\mu_{k,p}^{q}\left(q\!-\!1\right)},
 \label{eq:CDF3}
\end{align}
where $\bar{\gamma}_{{\tt th}}\triangleq \hat{\beta}_{llk}\left(
\alpha^2 -\alpha^{2}  C \gamma_{{\tt th}} \right)$, and where in
the third equality, we have used that the cumulative density
function of $X_k$ (Erlang variable) is
\begin{align}
F_{X_k}(x)&={\tt Pr}\left(X_k\leq x\right)\nn\\
&=1-\exp\left(-\frac{x}{\hat{\beta}_{llk}}\right)\sum_{t=0}^{N-K}\frac{1}{t!}\left(\frac{x}{\hat{\beta}_{llk}}\right)^t
\label{eq:CDF4}.
\end{align}
The last equality of \eqref{eq:CDF3} was derived after applying
the binomial expansion of $(y+1/p_r)^t$
and~\cite[Eq.~(3.351.1)]{GR:07:Book}.
\setcounter{eqnback}{\value{equation}} \setcounter{equation}{58}
\begin{figure*}[!t]
\begin{align}
 &\ddot{R}_{lk}\left( p_{r}, \alpha \right)=\frac{1}{\mathrm{ln}2} \EXs{X_k, Y_k}{\frac{\alpha^{2 }X_k[n-1]\left( \alpha^{4}X_k^{2}[n-1]+2\varsigma_k\left( 1+p_{r} \varsigma_k \right)^{2} \left( 1+\alpha^{2}p_{r} X_k[n-1] +p_{r} \varsigma_k\right)\right)}{\left( \alpha^{4}p_{r}^{2}\left(C+1  \right)X_k^{2}[n-1]+p Y_k[n]+1 \right)^{2}\left( \alpha^{4}p_{r}^{2}C X_k^{2}[n-1]+p_{r} Y_k[n]+1 \right)^{4}}},\label{eq: second_derivative}
\end{align}
\hrulefill
\end{figure*}
\setcounter{eqncnt}{\value{equation}}
\setcounter{equation}{\value{eqnback}}
\subsection{Proof of Theorem~\ref{Theo 3}} \label{sec:low snr}

The initial step for the derivation of the minimum transmit energy
per information bit is to cover the need for exact expressions
regarding the  derivatives of  ${R}_{lk}\left( p_{r}, \alpha
\right)$. In particular, this can be given by
\begin{align}\label{eq: first_derivative}
\!&\!\!\dot{R}_{lk}\left(\! p_{r}, \alpha \!\right)\!=\!
\frac{1}{\mathrm{ln}2}\nn\\
\!&\!\!\!\!\times\!\EXs{X_k, Y_k\!\!\!}{\!\!\frac{\alpha^{2 }X_k[n\!-\!1]\big/\!\!\left(\!
\alpha^{4}p_{r}^{2}C X_k^{2}[n\!-\!1]\!+\!p_{r} Y_k[n]\!+\!1
\!\right)\!\!}{\left(\! \alpha^{4}p_{r}^{2}\left(\!C\!+\!1
\!\right)\!X_k^{2}[n-1]\!+\!p_{r} Y_k[n]+1 \!\right)}}\!\!.
\end{align}
Easily, its value at $p_{r}=0$ is
\begin{align}\label{eq: first_derivative at zero}
 \dot{R}_{lk}\left( 0, \alpha \right)= \frac{1}{\mathrm{ln}2}\EXs{X_k}{\alpha^{2 }X_k[n-1]}.
\end{align}
Aknowledging that $X_k[n-1]$ is Erlang distributed, its
expectation can be written as
\begin{align}\label{eq: mean_value}
 \EXs{X_k}{X_k[n-1]}= \left( N-K+1 \right)\hat{\beta}_{llk}.
\end{align}
Substituting~\eqref{eq: mean_value} and~\eqref{eq:
first_derivative at zero} into~\eqref{eq: minimum energy}, we
lead to the desired result.

The second derivative of ${R}_{lk}\left( p_{r}, \alpha \right)$,
needed for the evaluation of the wideband slope, is given by
\eqref{eq: second_derivative} shown at the top of the previous
page, where $\varsigma_k \triangleq \alpha^{2}C X_k[n-1]+Y_k[n]$.
Hence, $\ddot{R}_{lk}\left( 0, \alpha \right)$ can be expressed
by
\addtocounter{equation}{1}
\begin{align}\label{eq: second_derivative
at zero} \ddot{R}_{lk}\left( 0, \alpha \right)
&=\frac{1}{\mathrm{ln}2}\mathbb{E}_{X_k, Y_k}\left\{
\alpha^{6}X_k^{3}[n-1]+2 \alpha^{4}C X_k^{2}[n-1]\right.
\nn\\
&\hspace{2.5cm}\left.+2 \alpha^{2}X_k[n-1]Y_k[n]\right\}.
\end{align}
The moments of $X_k[n-1]$ are obtained by means of the
corresponding derivatives of its moment generating function (MGF)
at zero $\mathrm{M}_{X_{k}}^{\left( n \right)}\left( 0 \right)$,
i.e., $\EXs{X_k}{X_k^{n}[n-1]}=\mathrm{M}_{X_{k}}^{\left( n
\right)}\left( 0 \right)$. Thus, having in mind that the MGF of
the Erlang distribution is
\begin{align}\label{eq: MGF}
\mathrm{M}_{X_{k}}\left(t  \right)=\frac{1}{\left( 1-
\hat{\beta}_{llk}t\right)^{N-K+1}},
\end{align}
we can obtain the required moments of $X_k[n-1]$ as
\begin{align}\label{eq: 1_MGF}
 \EXs{X_k}{X_k^{2}[n-1]}&=\mathrm{M}_{X_{k}}^{\left( 2 \right)}\left( 0 \right)\nn\\
 &=\frac{\Gamma\left( N-K+3 \right)}{\Gamma\left( N-K+1 \right)}\hat{\beta}_{llk}^{2}\\
 \EXs{X_k}{X_k^{3}[n-1]}&=\mathrm{M}_{X_{k}}^{\left( 3 \right)}\left( 0 \right)\nn\\
 &=\frac{\Gamma\left( N-K+4 \right)}{\Gamma\left( N-K+1 \right)}\hat{\beta}_{llk}^{3}.\label{eq: 2_MGF}
\end{align}

In addition, since $X_k[n-1]$ and $Y_k[n]$ are uncorrelated, we
have  $$\EXs{X_k,Y_{k}}{ X_k[n-1]Y_k[n]}=\EXs{X_k}{
X_k[n-1]}\EXs{Y_k}{Y_k[n]}.$$ In other words, it is necessary to
find the expectation of $Y_k[n]$. As aforementioned, the PDF of
$Y_k[n]$ obeys~\eqref{eq PDF1 3} and has expectation given by
definition as
\begin{align}
  \EXs{Y_k}{Y_k[n]}&=\int_{0}^{\infty}y\PDF{Y_k}{y}\mathrm{d}y\nn\\
    &=
        \sum_{p=1}^{\varrho\left( \mathbf{\mathcal{A}}_k\right)}
        \sum_{q=1}^{\tau_p \left( \mathbf{\mathcal{A}}_k\right)}\!\!
            \mathcal{X}_{p,q} \left( \mathbf{\mathcal{A}}_k\right)
            \frac{
                \mu_{k,p}^{-q}
            }{
                \left(
                    q-1
                \right)
                !
            }
           \int_{0}^{\infty}\!\! \!y^{q}
            e^{\frac{-y}{\mu_{k,p}}}\mathrm{d}y\nn\\
        &= \sum_{p=1}^{\varrho\left( \mathbf{\mathcal{A}}_k\right)}
        \sum_{q=1}^{\tau_p \left( \mathbf{\mathcal{A}}_k\right)}
            \mathcal{X}_{p,q} \left( \mathbf{\mathcal{A}}_k\right)
            \frac{
                \mu_{k,p}^{-q}q
            }{
                \left(
                    q-1
                \right)
                !
            },\label{eq: mean y}
\end{align}
where we have used~\cite[Eq.~(3.326.2)]{GR:07:Book} as well as the
identity $\Gamma\left( q+1 \right)=q!$. As a result,
$\ddot{R}_{lk}\left( 0, \alpha \right)$ follows by means
of~\eqref{eq: 1_MGF},~\eqref{eq: 2_MGF},~\eqref{eq: mean y}.
Finally, substitution of the~\eqref{eq: first_derivative at zero}
and~\eqref{eq: second_derivative at zero} into~\eqref{eq:
wideband_slope} yields the wideband slope.

\end{document}